\documentclass[conference]{IEEEtran}
\IEEEoverridecommandlockouts
\usepackage{cite}
\usepackage{url}
\usepackage{amsmath,amssymb,amsfonts}
\usepackage{algorithmic}
\usepackage{graphicx}
\usepackage{textcomp}
\usepackage{xcolor}
\usepackage{pifont}
\usepackage{multirow}
\usepackage{subfigure}
\usepackage{graphicx}
\usepackage{tikz}
\usepackage{textcomp}
\usepackage{color, soul}
\usepackage{booktabs}
\usepackage{comment}
\makeatletter
\def\UrlAlphabet{%
\do\a\do\b\do\c\do\d\do\e\do\f\do\g\do\h\do\i\do\j%
\do\k\do\l\do\m\do\n\do\o\do\p\do\q\do\r\do\s\do\t%
\do\u\do\v\do\w\do\x\do\y\do\z\do\A\do\B\do\C\do\D%
\do\E\do\F\do\G\do\H\do\I\do\J\do\K\do\L\do\M\do\N%
\do\O\do\P\do\Q\do\R\do\S\do\T\do\U\do\V\do\W\do\X%
\do\Y\do\Z}
\def\UrlDigits{\do\1\do\2\do\3\do\4\do\5\do\6\do\7\do\8\do\9\do\0}
\g@addto@macro{\UrlBreaks}{\UrlOrds}
\g@addto@macro{\UrlBreaks}{\UrlAlphabet}
\g@addto@macro{\UrlBreaks}{\UrlDigits}
\makeatother
\usepackage{todonotes}

\newcommand{\hcolor}{\color{black}}

\def\BibTeX{{\rm B\kern-.05em{\sc i\kern-.025em b}\kern-.08em
    T\kern-.1667em\lower.7ex\hbox{E}\kern-.125emX}}
\begin{document}

\title{Pagurus: Eliminating Cold Startup in Serverless Computing with Inter-Action Container Sharing 
}


\author{\IEEEauthorblockN{$^*$Zijun Li, $^{*\dagger}$Quan Chen, $^{*\dagger}$Minyi Guo}
\IEEEauthorblockA{$^*$\textit{Department of Computer Science and Engineering, Shanghai Jiao Tong University, China} }
\IEEEauthorblockA{$^\dagger$\textit{Shanghai Institute for Advanced Communication and Data Science, Shanghai Jiao Tong University, China} }
lzjzx1122@sjtu.edu.cn, \{chen-quan, guo-my\}@cs.sjtu.edu.cn}

\maketitle

\begin{abstract}
Serverless computing provides fine-grain resource sharing between Cloud tenants through containers.
Each function invocation (action) runs in an individual container.
When there is not an already started container for a user function, 
a new container has to be created for it. 
However, the long cold startup time of a container results in the long response latency of the action.
Our investigation shows that the containers for some user actions share most of the software packages. 
If an action that requires a new container can ``borrow'' a similar warm container from other actions, the long cold startup can be eliminated. 
Based on the above finding, we propose Pagurus, a runtime container management system for eliminating the cold startup in serverless computing. 
Pagurus is comprised of an inter-action container scheduler and an intra-action container scheduler for each action. 
The inter-action container scheduler schedules shared containers among actions. 
The intra-action container scheduler deals with the management of the container lifecycle.
Our experimental results show that Pagurus effectively eliminates the time-consuming container cold startup. An action may start to run in 10ms with Pagurus, even if there is not warm container for it.
\end{abstract}


\section{Introduction}
Adopting serverless computing,
 Cloud tenants submit functions directly without renting virtual machines of different specifications, Cloud vendors schedule the tenants' functions automatically.
For the high maintainability and testability, most hyperscalers now provide serverless computing services (such as Amazon Lambda~\cite{lambda}, Google Cloud Function~\cite{googlefunction}, Microsoft Azure Functions~\cite{azurefunction}, and Alibaba Function Compute~\cite{Alicloud}).
Serverless computing is perfect for the Internet services that have unstable query loads, since the tenants are charged for each query execution, instead of the long term renting.

We use the terminology in Apache OpenWhisk~\cite{openwhisk}, an event-driven serverless computing platform. An {\it action} represents the invocation of a user function. Whenever an action is received, serverless computing runs the action using either a newly-launched container or a running warm container. 
The warm containers keep serving queries ({\it warm startup}) until timeout to be recycled. 
While containers of different actions rely on various software packages, the containers are not shared.
If there is not a warm container for an action, a new cold container needs to be started for it. The long latency of booting containers, as well as the software environment and code initialization, restricts the performance of serverless computing~\cite{DBLP:conf/icdcsw/McGrathB17,wang2018peeking,serverless_report,DBLP:conf/nsdi/PuVS19,DBLP:conf/micro/ShahradBW19}.


While some actions suffer from the long container cold startup time, we observe that current serverless computing systems may launch too many containers for some other actions. For instance, they may launch too many containers for Internet services with diurnal load patterns~\cite{barroso2003web,dean2013tail} (the low load is less than 30\% of the peak load) at the peak load. We also observe that there are idle containers for some actions, even if loads of these actions are stable. While these warm containers are idle and wasting the system resources, they are not used by any other action because the containers for different actions install different software packages. 

With the development of micro-service architecture, the actions tend to use popular and common library package. For instance, by extracting likely dependencies in the projects on packages in the popular Python Package Index (PyPI) repository, 36\% of imports are of 20 popular packages\cite{sock}. {Actions tend to use similar software packages.} In this scenario, if an action that requires cold container startup is able to utilize the idle containers of other actions, the cold startup is turned into a warm startup, its end-to-end latency can be greatly reduced. 

There are three main challenges in achieving the above purpose. As for the first challenge, the loads of the actions are not stable~\cite{DBLP:journals/corr/abs-1810-09679}. It is difficult to determine whether an action can safely lend an idle container to other actions, without affecting its own Quality-of-Service (QoS).
Secondly, existing serverless computing systems do not support the ``borrow'' operation. Container sharing is not allowed.
Thirdly, multiple renters and lenders coexist. It is non-trivial to design an efficient container sharing strategy among actions that minimizes the number of the cold startup.


To tackle the above challenges, we propose {\bf Pagurus}, a container management system that reduces container cold startup through adaptive inter-action container sharing. In Pagurus, the containers are classified into {\it lender containers}, {\it executant containers} and {\it renter containers}. The executant containers can only be used by the owner action itself. 
The lender containers can be lent to other actions, and will turn into renter containers. 
Pagurus proposes an enhanced container component that enables the container sharing between multiple actions and guarantees the security of users' code. For each action, an {\it intra-action container scheduler} is adopted to manage its executant containers, the renter containers borrowed from other actions, and the lender containers to be lent to other actions. The whole serverless computing system adopts an {\it inter-action container scheduler} to schedule the containers between the actions and handles the proactive re-packing based on the package similarity of the actions and their workloads.

The main contributions of this paper are as follows.
\begin{itemize}
  \item {\bf The container enhancement that enables the sharing.} We enhance the container design to support the runtime package re-packing. It enables the container sharing between different actions.
  \item {\bf The design of a similarity-based container re-packing policy.} While the containers for different actions install different software packages, we analyze the similarities of the actions, and minimize the number of packages installed for container sharing.
  \item {\bf The design of an efficient inter-action container sharing mechanism.} This mechanism divides the containers into three types, based on which Pagurus manages them in different ways and enables efficient inter-action container sharing. 

\end{itemize}

Through the adaptive inter-action container sharing, Pagurus greatly reduces the possibility that an action suffers from cold container startup. Pagurus is also able to be integrated with prior work on reducing the container cold startup time to minimize the overhead of serverless computing in all the cases.


\section{Background and Related work}
\label{sec:related}
\subsection{Background}
In serverless computing, container act as a lightweight virtualization that creates multiple isolated user-space instances for actions. The serverless platform uses containers to encapsulate and execute the queries.

\begin{figure}
  \centerline{\includegraphics[width=.9\columnwidth]{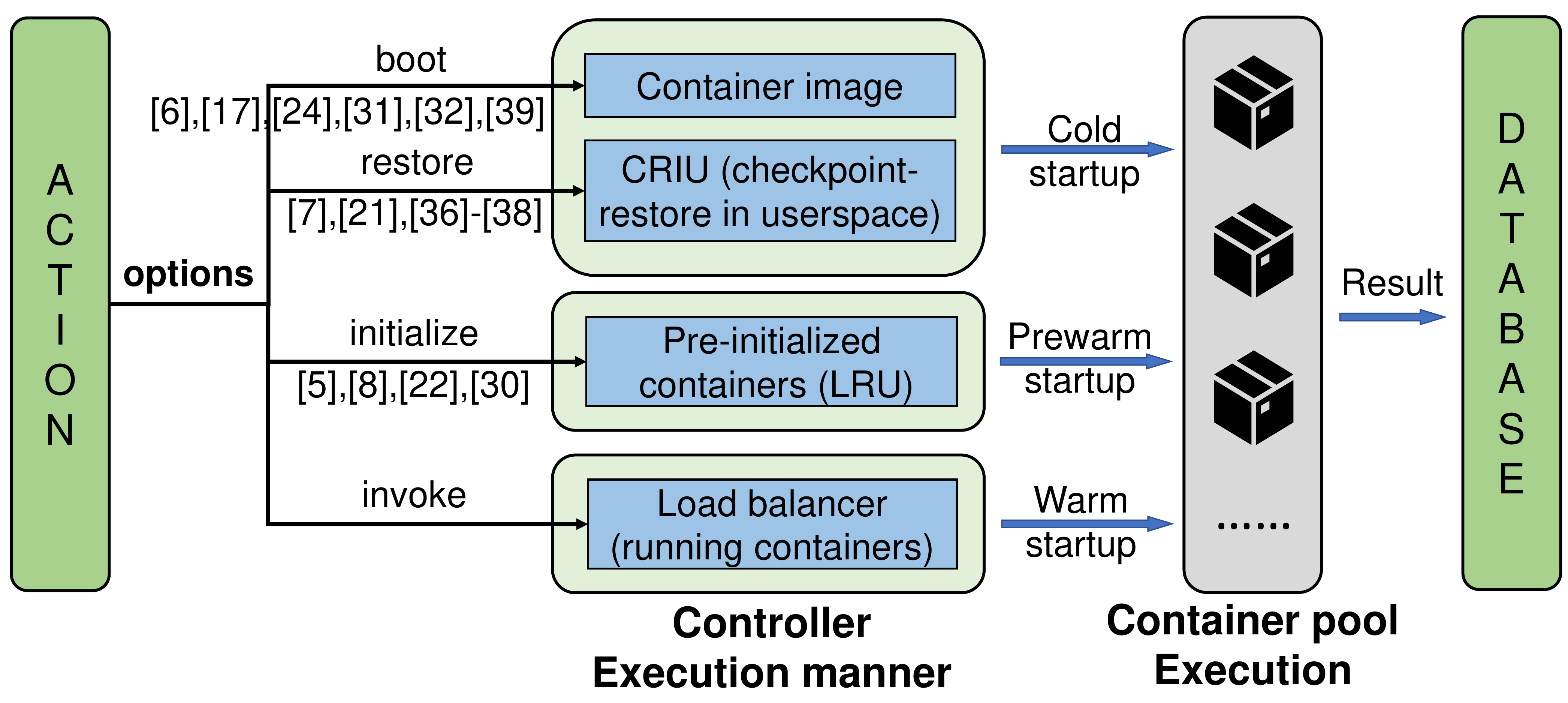}}
  \caption{\label{fig:serverless} Four possibilities to start an action in a serverless computing system. 
  }
\end{figure}

Figure~\ref{fig:serverless} shows the way that a user action is scheduled to run with serverless computing. As shown in the figure, a container must be booted/restored/initialized/invoked to host the action. If an action is invoked for the first time or there is no alive container for this action, the serverless system encapsulates it and starts up a new container, initializes software environment, loads application-specific code~\cite{Knative}, and runs the function. All these steps make up a {\it cold startup}~\cite{jonas2019cloud,baldini2017serverless}, and may take several seconds. The container cold startup significantly increases the end-to-end latency of user queries~\cite{wang2018peeking,serverless_report,DBLP:conf/nsdi/PuVS19,DBLP:conf/micro/ShahradBW19}, especially the processing of a single query is often short (hundreds of milliseconds) for Internet services.

To deal with this problem, researchers also proposed to use CRIU (checkpoint-restore in user-space)~\cite{CRIU,DBLP:conf/eurosys/WangHW19,DBLP:conf/sosp/VrableMCMVSVS05,Venkatesh2019fast} technique that restores container images from checkpoints to reduce the cold startup time. However, it still incurs the long end-to-end latency~\cite{du2020catalyzer}. Another approach called {\it prewarm startup} adopted by OpenWhisk spawns stem cell containers that are already initialized with the software environment in advance. Though it skips the container startup and users only need to perform application-specific code initialization~\cite{openwhisk,openwhisk_doc,DBLP:conf/fast/HarterSLAA16,sock}, its pre-loaded libraries can either make the image size too large~\cite{DBLP:conf/fast/HarterSLAA16,DBLP:conf/pppj/2018}, or cause long startup latency for the prewarm container~\cite{sock,DBLP:conf/cloud/Brewer15}.


If a container for a type of action is alive (just complete the previous invocation), a new action query of the same type can be directly executed in the running container ({\it warm startup}). Warm startup eliminates the container booting and initialization, and these warm containers keep serving actions to achieve better end-to-end latency~\cite{DBLP:conf/icdcsw/McGrathB17}. However, warm startups are not always possible because these warm containers may be recycled before cold startup happens~\cite{jonas2019cloud,baldini2017serverless}.

\subsection{Related Work}
There are already lots of prior work on reducing the container startup latency to improve the performance of serverless computing~\cite{DBLP:journals/usenix-login/HendricksonSOHV16,akkus2018sand,du2020catalyzer,DBLP:conf/asplos/ShenSSBDRW19}. SAND~\cite{akkus2018sand} separates different applications from each other via containers, while allowing functions of one application to run in the same container by different processes.
X-container~\cite{DBLP:conf/asplos/ShenSSBDRW19} has been proposed as a new security paradigm for isolating cloud-native containers to achieve higher throughput. Catalyzer~\cite{du2020catalyzer} adopts the design that utilizes the technology of CRIU with on-demand recovery. Hendrickson~\cite{DBLP:journals/usenix-login/HendricksonSOHV16} also proposes OpenLambda to deal with the long function startup latency and locality consideration.

Slacker~\cite{DBLP:conf/fast/HarterSLAA16} and SOCK~\cite{sock} share the similar idea that containers are launched by generalizing zygote initialization to reduce the startup latency.
To achieve function isolation, Unikernels~\cite{DBLP:journals/cacm/MadhavapeddyS14,DBLP:conf/hotos/KollerW17} can achieve less latency and better throughput via bypassing the kernel with unikernels in serverless environments.
McGrath~\cite{DBLP:conf/icdcsw/McGrathB17} proposed to reuse containers and create containers by introducing a queuing scheme with workers collecting the availability in different queues.

Existing works mainly focus on seeking more lightweight virtualization technologies to pursue lower overhead, or to reduce the container startup time for one kind of action. Our work tries to make different actions work collaboratively to alleviate the container cold startup problem. Furthermore, Pagurus can be combined with different container technologies to achieve less cold startup latency.

\section{Motivation}
\label{moti}

\subsection{Experimental Setup}
\label{sec:setup}
In this investigation, we use Apache OpenWhisk~\cite{openwhisk} with local cache as the representative serverless computing platform, FunctionBench~\cite{functionbench} and Faas-Profiler~\cite{DBLP:conf/micro/ShahradBW19} as the benchmarks. The experiment setup is based on a 2-node cluster where the nodes are connected with a 25Gb/s Ethernet switch. In the 2-node experimental cluster, we use one node to perform the computing, and one node to generate the queries for execution. 
Table~\ref{tab:environment} shows the hardware and software configurations of each node. 
We use representative serverless computing benchmark suites FunctionBench~\cite{functionbench} and Faas-profiler~\cite{DBLP:conf/micro/ShahradBW19} and the used benchmark workloads are shown in Table~\ref{tab:benchmarks}.

\newcommand{\tabincell}[2]{\begin{tabular}{@{}#1@{}}#2\end{tabular}}
\begin{table}
\centering
    \caption{\label{tab:environment}Hardware and software setups}
    \scriptsize
    \begin{tabular}{c|c}
        \hline
        & Configuration\\
        \hline
        Node & \tabincell{c}{CPU: Intel Xeon Platinum 8163@2.50GHz\\Cores: 40, L3 shared cache: 32MB\\DRAM: 256GB,  Disk: NVME SSD \\
        Network Interface Card (NIC): 25Gb/s} \\
        \hline
        Network & 25Gb/s Ethernet Switch \\
        \hline
        Software&\tabincell{c}{Nginx version: nginx/1.10.3  Database: Apache/couchdb:2.3 \\  Container runtime: Python-3.7.0, Linux with kernel 4.15.0\\ Docker server and client version: 19.03 \\ Docker runc version: 1.0.0-rc10  Docker containerd version: 1.2.13 \ \\Memory and Timeout of serverless containers: 256MB, 60s \\ Operating system: Linux with kernel 3.10.0} \\
        \hline
    \end{tabular}
\end{table}

\begin{table}
\vspace{-2mm}
  \centering{}
  \caption{\label{tab:benchmarks}Benchmarks used in this paper}
  \scriptsize
  \begin{tabular}{l|l|l}
    \hline
   Benchmark & Workloads & Description\\
    \hline
    & {\it dd} & Convert and copy a file.\\
    & {\it float\_operation(fop)} & Float operations (sin, cos and sqrt).\\
    & {\it cloud\_storage (clou)} & Cloud storage service.\\
    & {\it mapreduce (mr)} & MapReduce wordcount workload.\\
  FunctionBench  & {\it video\_processing (vid)} & Video processing.\\
    & {\it linpack (lp)} & Solve linear equations $Ax = b$.\\
    & {\it matmul (mm)} & Matrix multiplication.\\
    & {\it k-means (kms)} & Model training of k-means.\\
  \hline
    & {\it image\_resize (img)} & Resizes an image to several icons.\\
    Faas-profiler &  {\it couchdb (cdb)} & Json\_dump from Coucdb files.\\
    & {\it markdown2htmll (md)} & Renders Markdown text to HTML.\\
    \hline
  \end{tabular}
\end{table}

\subsection{Breakdown of the End-to-end Latency}
The end-to-end latency of processing a user's query seriously affects the user experience. 
We make an investigation and break down the end-to-end latencies of the benchmarks for serverless computing.

\begin{figure}
\centerline{\includegraphics[width=.9\columnwidth]{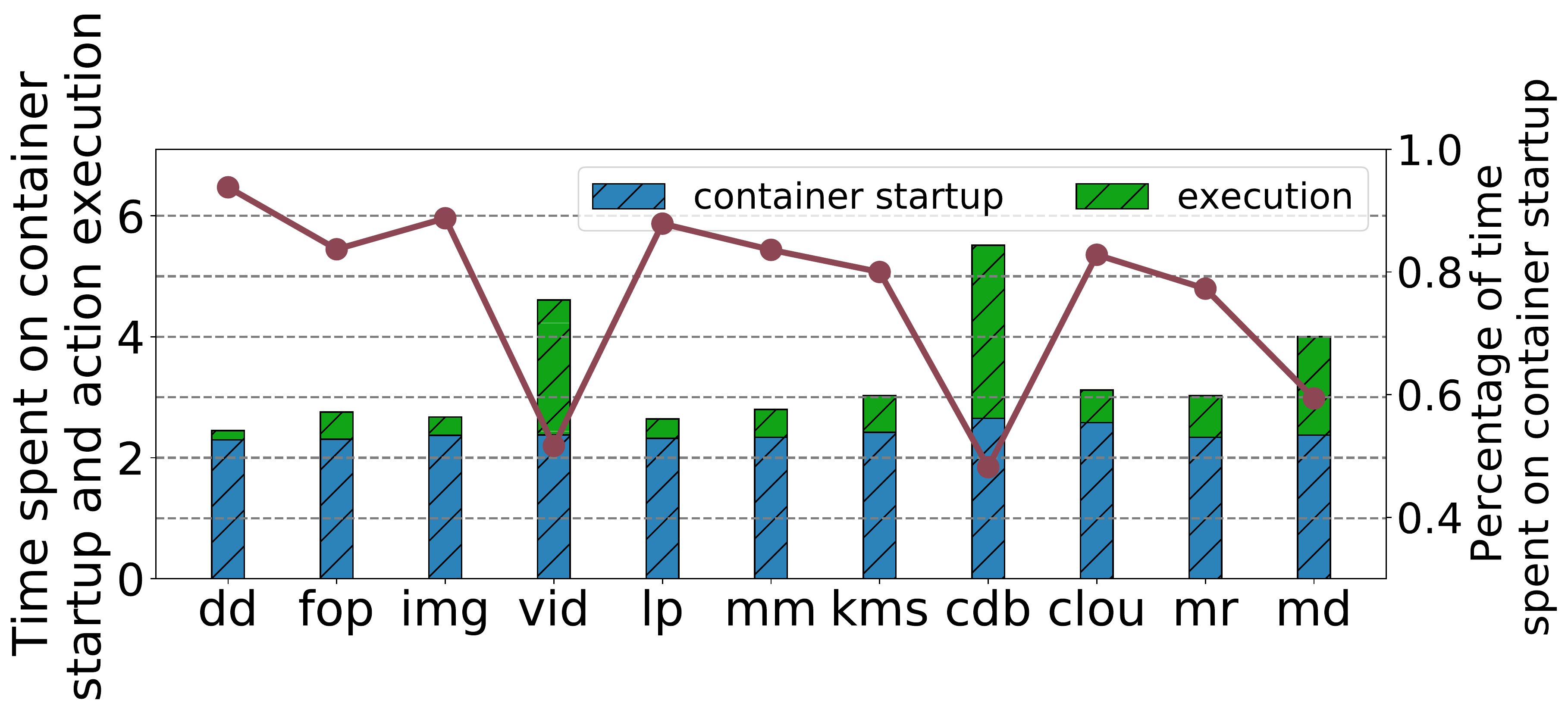}}
\caption{\label{fig:coldstart} The distribution and percentages of time spent on the cold container startup and action execution in the end-to-end latencies of the representative benchmarks.}
\vspace{-3mm}
\end{figure}

In a serverless computing system based on container technology, cold container startup happens when there are no idle containers exist, and a user query is received. In this scenario, the system creates a new container to serve the query.
 For an action, the cold container startup includes operations like initializing the customized execution environment. 
Traditionally, the cold startup overhead includes the container startup, the software environment of the function initialization, and application-specific code initialization. These operations may incur significant extra latency.
 Figure~\ref{fig:coldstart} shows the percentages of the time spent on the cold container startup in the end-to-end latencies of the benchmarks.
 
Observed from Figure~\ref{fig:coldstart}, 
the cold startup overhead increases the end-to-end latencies of the benchmarks. 
In general, the container cold startup time is relatively stable. 
In the best case, the cold container startup still takes 48.2\% of the end-to-end latency ({\it cdb}).
While in the worst case, the cold container startup takes 93.8\% of the end-to-end latency ({\it dd}).

If we can eliminate the container cold startup, the end-to-end latencies of the applications with serverless computing can be greatly reduced. To this end, Cloud vendors~\cite{googlefunction,Alicloud,azurefunction,lambda,gVisor,Firecracker}, as well as recent works~\cite{DBLP:journals/usenix-login/HendricksonSOHV16,baldini2017serverless,DBLP:conf/sosp/MancoLSMKSYRH17,DBLP:conf/usenix/ThalheimBFK18} have focused more on reducing the container startup time as we discussed in Section~\ref{sec:related}.

{\hcolor Even if the container cold startup time is reduced to about 40ms in the best case~\cite{du2020catalyzer}, the cold startup still takes longer time than the case that the query can directly get a warm container ($<$10ms)~\cite{openwhisk,DBLP:conf/micro/ShahradBW19}. 
The increasing usage of high-level language like Python, can make the cold startup even more expensive\cite{akkus2018sand,DBLP:conf/micro/ShahradBW19}. 
For the latency-sensitive applications that have millisecond-level latency targets, such as Internet services, the 30ms already results in poor user experience.}

\subsection{Existence of Redundant Warm Containers}
\begin{figure}[!t]
	\centering
	\subfigure[The number of containers launched in OpenWhisk]{
		\includegraphics[width=.42\textwidth]{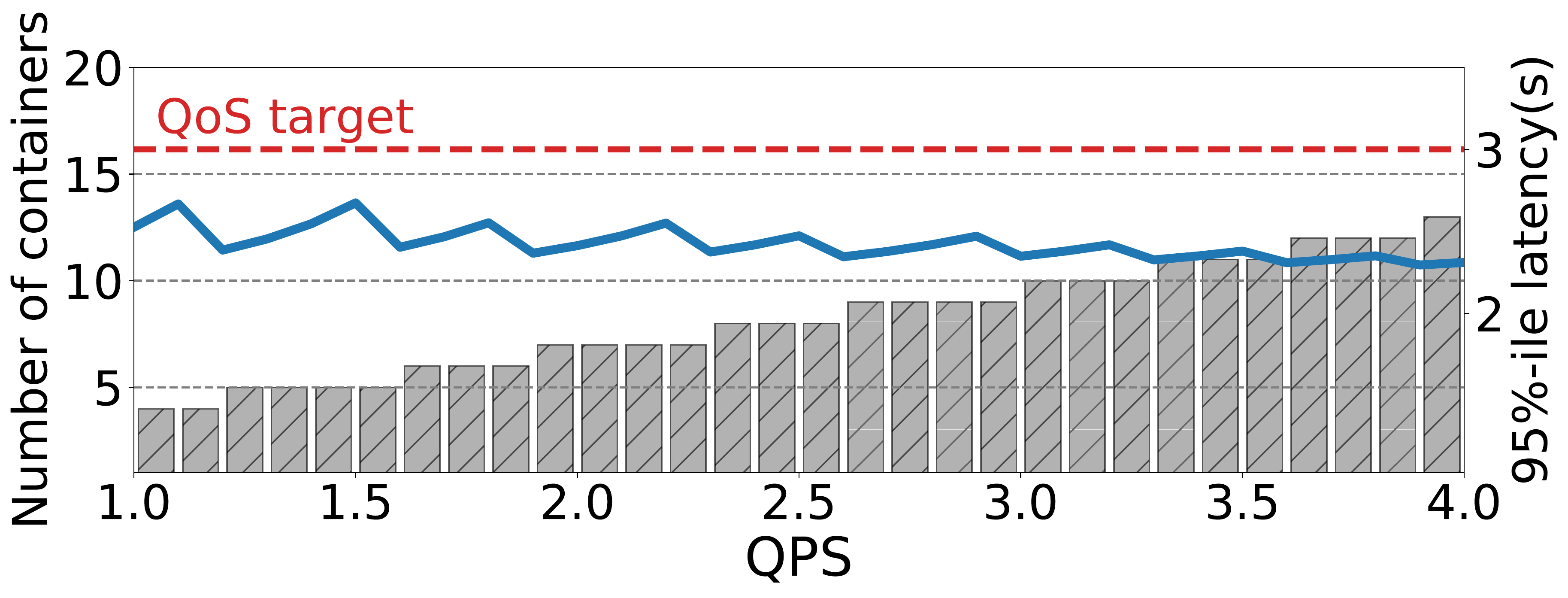}
	}
	\subfigure[The number of actually needed containers]{
		\includegraphics[width=.42\textwidth]{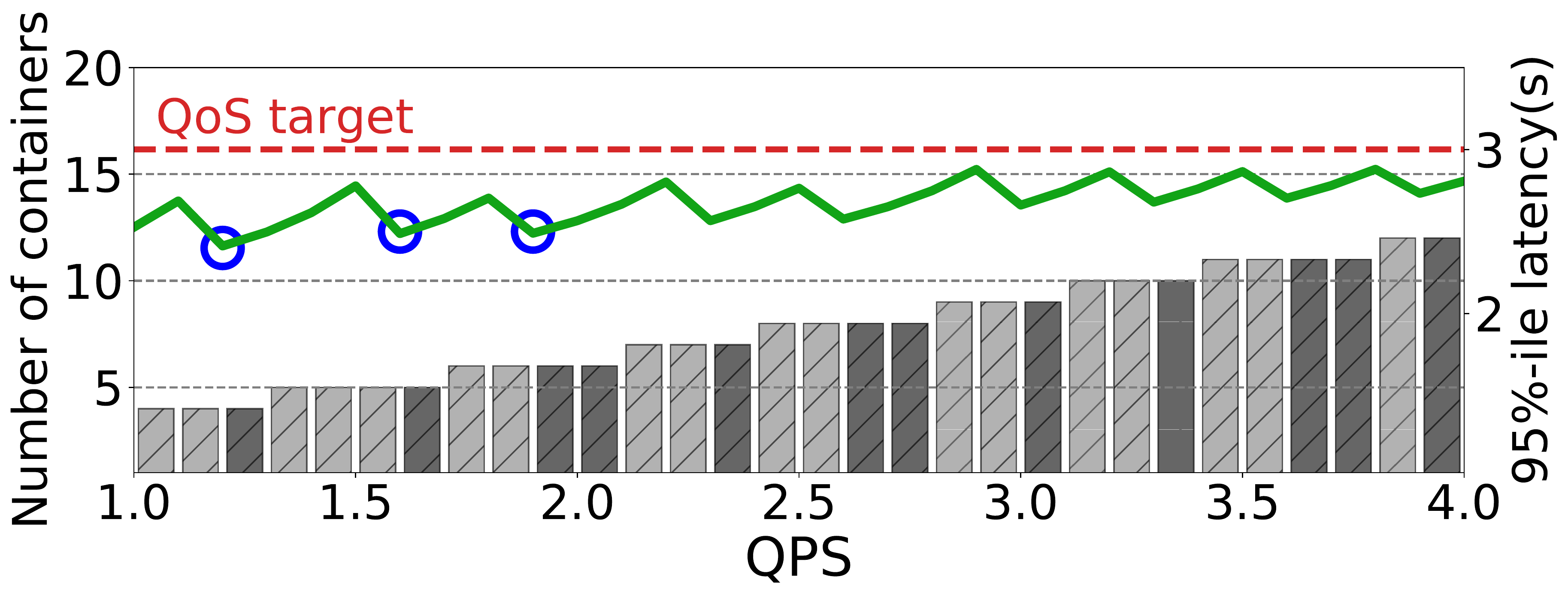}
	}
	\caption{The number of containers launched (a) and required (b) to ensure the 95\%-ile latency target of an example benchmark {\it vid}.}
	\label{fig:container}
	\vspace{-3mm}
\end{figure}

An important feature of serverless computing is elasticity. By current container startup strategy, the containers are started up upon queries waiting in the queue, and will be recycled soon when there is no query for a certain period (e.g., 60 seconds in OpenWhisk).
Therefore, whenever the running containers fail to catch up with the queries waiting in the queue, a new container must get started and experience cold startup. 
In a large-scale serverless computing platform, there may coexist a large number of actions from different users. If some actions have redundant warm containers,
we envision that it is possible to 
reuse these redundant warm containers to eliminate the cold startup.
To this end, we manually schedule the container startup process to reuse the redundant warm containers, and check the number of containers launched and the QoS in the metric of end-to-end latency. Figure~\ref{fig:container} reports our investigation results.
Our investigation proves that our assumption holds. 

Taking the benchmark {\it vid} as an example, Figure~\ref{fig:container}(a) and Figure~\ref{fig:container}(b) show the number of containers launched and the 95\%-ile end-to-end latencies of the queries in various loads, for OpenWhisk and our manual scheduling, respectively.
In the figure, the $x$-axis represents the query workload (Query-per-Second, QPS); the bars show the number of warm containers (corresponding to the left $y$-axis), and the line shows the 95\%-ile latency (corresponding right $y$-axis). 

Observed from Figure~\ref{fig:container}(a), the 95\%-ile latency of the benchmark shows the pattern of cyclic variation. 
This is because a little increase in the QPS at the saturation point will bring in a new container startup, thanks to which, lower overall latency will be achieved. But further increasing the QPS will result in higher overall latency.
While, there is still headroom between the 95\%-ile latency and the QoS target, it is possible to use fewer containers without violating the QoS requirement. This is exactly what we do in the manual scheduling.
As shown in Figure~\ref{fig:container}(b), in some cases (the bars in black), we can safely reduce at least one container while still achieving the 95\%-ile latency target. 
It proves that some applications in the serverless computing platform do have some warm idle containers during execution. 
In addition, from Figure \ref{fig:container}(b), we also observe that the idle container usually appears in a minimum turning point of the latency (represented by blue circle in the figure). It is because that the minimum turning point usually comes with a new container startup to deal with the increasing queries in the waiting queue.
We also find that it is a common phenomenon based on Openwhisk. Besides $vid$, the other benchmarks also produce the similar results.
On the other hand, it can be easily anticipated that there will be even more idle containers in the case when the query workload drops suddenly.
While, it has been widely witnessed and proved that real Internet services are with diurnal load pattern~\cite{barroso2003web,dean2013tail}. Definitely, this will bring in more idle containers potentially for reuse.

\subsection{Challenges in Reusing Containers and Ensuring QoS}
Based on the above analysis and investigation, there is an opportunity to leverage the warm containers of some actions to eliminate the cold container startup of other actions. 
While the actions may require different software packages and execution environment, an action's container is not able to be used by another container. An action's container has to be re-packed before it can be used by other actions. The re-packing operation may take a relatively long time. 

By extracting the packages from the benchmark suite FunctionBench~\cite{functionbench}, we also find that 16.7\% of the benchmarks import $pandas$ and $sklearn$, and some commom libraries like $numpy$ are even shared by 22.2\% of them. This finding indicates that, different actions tend to share a high proportion of packages with others. Therefore, it is possible to build a shared container image, allowing several actions to run without installing extra packages. Even if in some cases the similarity of libraries between different actions is not high, we can use specific designed algorithms to build some connection between them.
However, it is not an easy task to build a shared container image as there are still several challenges, such as 



\begin{itemize}
    \item \textbf{Actions are not able to share containers.} While the containers of different actions pack different software packages, the containers of an action cannot be reused by other actions by default. 
    
    \item \textbf{Container reuse brings extra time overhead.} To allow other actions to run in an action's container, the reused container is supposed to install extra packages. Inappropriate package installation brings large time overhead that negates the latency reduction from the elimination of the cold container startup.
    \item \textbf{Security concerns about inter-action container sharing.} When containers are shared between different actions, isolation is weakened. While, the security and privacy of the actions must be ensured.
    \item \textbf{Inter-action container reuse brings extra schedule complexity.} While multiple actions are active concurrently, an efficient mechanism has to be proposed to manage the container lend and rent between the actions.
\end{itemize}

\section{Design of Pagurus}
\label{design}
To tackle the above challenges, we propose {\bf Pagurus}, a runtime container management system for eliminating the cold container startup through inter-action container sharing.
\label{sec:multi-nodes}

\begin{figure}
\centerline{\includegraphics[width=\columnwidth]{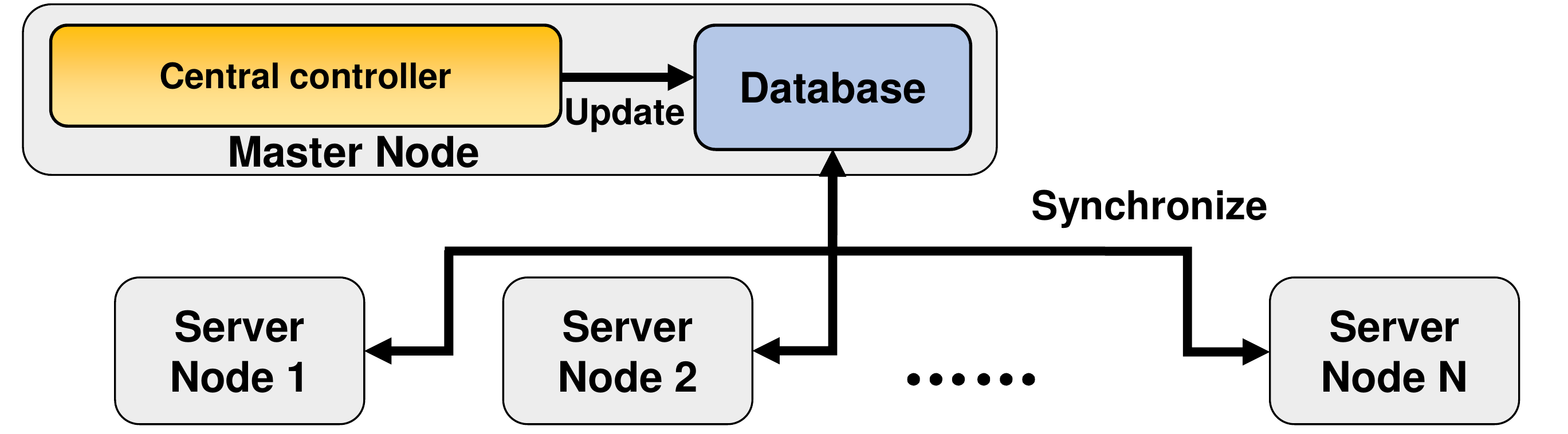}}
\vspace{-2mm}
\caption{\label{fig:multi-nodes}The Master-Slave design of a serverless computing system.}
\vspace{-3mm}
\end{figure}

For a traditional serverless computing system, the design of  distributed deployment is usually implemented in two ways. One is to divide the nodes into master and slave nodes. Load balancing is realized by the central controller in the master node, and the data in the slave node (server node) is synchronized by the database, as shown in Figure~\ref{fig:multi-nodes}. However, according to the previous studies, the network bandwidth between the server nodes and the database is usually the bottleneck of the serverless computing, and such master-slave design is also unpractical~\cite{jonas2019cloud,sock,DBLP:journals/usenix-login/HendricksonSOHV16}.

\begin{figure}
\centerline{\includegraphics[width=\columnwidth]{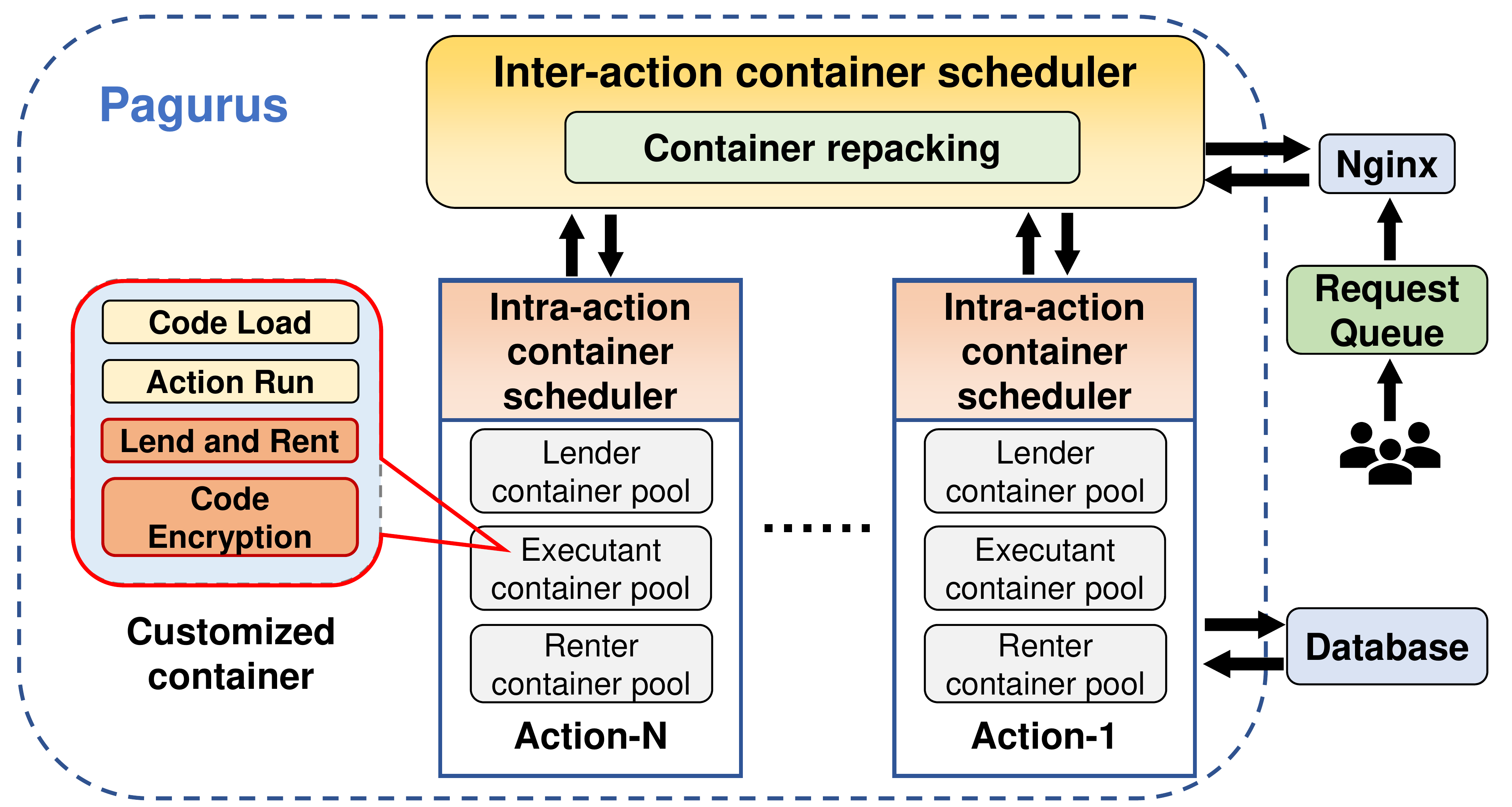}}
\vspace{-2mm}
\caption{\label{fig:pagurus}Design of Pagurus.}
\vspace{-3mm}
\end{figure}

Therefore, we design Pagurus using the single node management, where all nodes communicate with each other to maintain and update local databases and files. By such means, the performance is only relevant to the computing power of the server node. Figure~\ref{fig:pagurus} shows the design overview of Pagurus.
For the management of shared containers between actions,  an {\it inter-action container scheduler} is introduced.
It is also responsible for the re-packing of the containers at runtime when necessary.
For each action, there is an {\it intra-action container scheduler} responsible for coordinating three container pools, i.e., an executant container pool, a lender container pool, and a renter container pool. 
Whenever a container experiences the cold startup, it is added to the executant container pool by default, and will keep intact provided that it is recognized as a warm container.
When there is no query requesting the container for a certain period, the container will be identified as idle and moved to the lender container pool for possible cross-action reuse.
The renter container pool reserves the containers rent from the other actions' lender container pool.
Once a cold startup is about to occur on an action, Pagurus allows it to first check whether it is possible to reuse an existing lender container from another action to avoid cold startup. It should be noticed that the lender actions can also get renter containers whenever it is re-packed by others.

To enable the container sharing, we customize the container structure of Pagurus with four modules, i.e., {\it code\ load}, {\it action\ run},  {\it lend\ and\ rent} and {\it code\ encryption}, as shown in Figure \ref{fig:pagurus}.
{\it Code\ load} and {\it action\ run} are the same as current serverless computing platform, responsible for code loading from database and execution monitoring when invoking actions, respectively.
 {\it Lend\ and\ rent} and {\it code\ encryption} are specially introduced in Pagurus for container sharing and safety guarantee, respectively.
 {\it Lend\ and\ rent} essentially consists of  {\it lend} and {\it rent} functions.
{\it Lend} helps the container transferred from executant container to lender container. {\it Rent} helps renter inherit containers and properties from lender container. 
In addition, {\it code\ encryption} module guarantees the container security during code reload. 

Container re-packing and container life-cycle management play critical roles in Pagurus. We will discuss the two issues in the following two sections, respectively, with special emphasis on answering the following questions:
\begin{itemize}
    \item On what conditions, we can identify a container in the executant container pool as idle and transfer it to lender container pool?
    \item For a set of renter candidates, how to select the appropriate renter container in the consideration of runtime performance efficiency in the metric of end-to-end latency of renter containers?
    \item How to guarantee the security of lender and renter without exposing both sides' code and data?
\end{itemize}

\section{Image Re-packing}

\subsection{Idle Container Identification}
\label{sec:mmn}
Idle container identification and lender container generation are two essential functions in the intra-action container scheduler. To realize container sharing, it is first important to identify the containers of an action that can be lent, i.e., idle container.
In the serverless system, the queries are executed by the running containers. If the total capacity of the containers exceeds the one required to ensure the QoS of the queries during a period, idle container arises.
As we have known, whether a container can be judged as idle or not, depends on the query workload, the processing power, and the desired QoS.
Therefore, we first describe the query processing logic of serverless computing into a producer-consumer problem, and analyze it via queuing theory, which has been widely applied in  communication systems, computations and storage systems.


Without loss of generality, the query arrival process to a container can be described as Poisson process with exponentially distributed interval averaged as $\lambda$. The query processing time follows an exponential distribution with average value $\mu$, which is independent to the task arrival. The queries are fairly allocated among the containers. Thus, we can apply $M/M/n$ model~\cite{gautam2012analysis} to analyze the processing process.

When the traffic density $\rho = \frac{\lambda}{n\mu} <1$, the system is in a stable state.
In this case, we can derive the stable distribution $\pi_k$ that there are $k$ queries in the waiting queue as
\begin{equation}
\label{eq:1}
\scriptsize
\pi_k=\left\{
\begin{aligned}
\frac{(n\rho)^k\pi_0}{k!} , k= 1, 2, \dots, n-1\\
\frac{n^n\rho^k\pi_0}{n!} , k= n, n+1, \dots,
\end{aligned}
\right.
\end{equation}
where $\pi_0 = [\sum^{n-1}_{k=0}\frac{(n\rho)^k}{k!}+\frac{(n\rho)^n}{n!(1-\rho)}]^{-1}$. For brevity, the detailed analysis is omitted.
Then, we can further derive the average waiting time $W$ under the stable state (i.e.,  $\rho < 1 $). 
No query will be in the waiting queue if the number of queries is less than the number of containers, i.e.,  $P\{W=0\}=p\{X < n\}$.
Thus, we can derive the waiting time (i.e., the time spent in the waiting queue) distribution as
\begin{equation}
\label{eq:2}
\scriptsize
\begin{aligned}
P\{W\!=\!0\}& = p\{X\!<\!n\} = 1\!-\!p\{X \geq n\}\!= \! 1\!-\! \sum^{\infty}_{k=n}\pi_k  \\
& =1- \sum^{\infty}_{k=n}\frac{n^n\rho^k}{n!}\pi_0 =1- \frac{\pi_n}{1-\rho},
\end{aligned}
\end{equation}
and
\begin{equation}
\label{eq:3}
\scriptsize
\begin{aligned}
P\{0\!<\!W\!<\!t\} &= \sum^{\infty}_{k=n}P\{w\!<\!W\!<\!t|X\!=\!k\}P\{X\!=\!k\} \\
& = \sum^{\infty}_{k=n}\pi_{k}\int^t_0\frac{(n\mu)^{k-n+1}x^{k-n}}{\Gamma(k-n+1)}e^{-n\mu x}dx \\
& = \int^t_0\pi_n\sum^{\infty}_{k=n}\frac{(n\mu x \rho)^{k-n}}{(k-n)!}n\mu e^{-n\mu x}dx \\
& = \frac{\pi_n}{1-\rho}[1-e^{-n\mu(1-\rho)t}].
\end{aligned}
\end{equation}

Summing up \eqref{eq:2} and \eqref{eq:3}, we obtain the general waiting time distribution $F_w(t)$ as
\begin{equation}
\scriptsize
\begin{aligned}
\label{eq:4}
F_w(t) &  = P\{W = 0\} + P\{0<W<t\} \\
& = 1-\frac{\pi_n}{1-\rho}e^{-n\mu(1-\rho)t} , t>0 .
\end{aligned}
\end{equation}

Let $T_D$ represent the QoS target and define $r_{real}(n)$ as the $r_{real}$-ile latency of an action when there are $n$ containers, and $r_{req}$ as the $r_{req}$-ile latency requested by an action. So when the waiting time $t$ is set as the maximum waiting time $T_D-\frac{1}{\mu}$, $F_w(t)-r_{req} \ge 0 $ will represent whether the QoS requirement can be satisfied.
Thus, we can derive the discriminant function to determine whether the idle container of an action exists as
\begin{equation}
\label{eq:5}
\scriptsize
\left\{\begin{array}{l}
r_{real}(n) - r_{req} \ge 0 \\
\hat{f}(n-1) = 1 - r_{req} -\frac{\pi_n}{1-\rho}e^{-(n-1)\mu(1-\rho)(T_D-\frac{1}{\mu})} \ge 0 .
\end{array}\right.
\end{equation}

Both criteria in  \eqref{eq:5} need to be satisfied to identify an idle container. It is necessary that $r_{real}(n) \ge r_{req}$, otherwise the current QoS of the action can not be satisfied when $n$ containers running in serverless platform, and the action suffers QoS violation due to cold startup.
Upon QoS satisfaction, we further try to evaluate whether it is possible to remove one container from the executant container pool by checking the achievable QoS after the removal.
If $\hat{f}(n-1) \ge 0$,  $n-1$ containers are already enough to satisfy the QoS requirement, and an idle container will be removed by its intra-action container scheduler. Meanwhile, the inter-action container scheduler will re-pack the lender image for it.

Thus, the intra-action container scheduler of an action can apply the criteria in~\eqref{eq:5} to identify the idle containers for possible reuse by the other actions, and send the corresponding container information to the inter-action container scheduler for lender container image re-packing, as will be discussed in the next subsection.

\subsection{Similarity-based Re-packing}
\label{sec:steps}
Re-packing refers to adding extra dependent libraries to maximize the possibility of reuse by the other actions as different actions usually ask for containers with different libraries. Intuitively, we may add arbitrary more libraries to build a lender container for the maximal reuse. However, this will result in an extremely large container, leading to high overhead. Fortunately, we notice that different actions also share some libraries with different degrees. This motivates us to design a similarity-based container re-packing in Pagurus.

The inter-action container scheduler analyzes the software environment of each action, and re-packs the lender container image for each running intra-action container scheduler by checking the similarities between lender actions and other actions.
To filter out actions similar to a lender action, we apply collaborative filtering and adopt Nearest Neighbor Search (NNS) to calculate the similarity between two actions. Cosine-based Similarity is a well-known similarity algorithm which is traditionally used in users' interests recommendation. We define the action-$L$ as the actions that require additional libraries, and action-$NL$ as actions without additional libraries. The inter-action container scheduler generates the lender containers images by re-packing the similar images in the following steps.

\begin{itemize}
    \item {\bf Collect information about all actions}. All the information about their libraries will be recorded, including the name and the version of libraries. For each action, by the user's $Dockerfile$, additional installed libraries can be recorded in the format $\{L_n\} = \{"lib_1":"version","lib_2":"version"\}$. In some cases when users do not declare the version of libraries, the inter-action container scheduler will take the latest version as default.
    But it will bring in the hazard of libraries version contradiction. For example, if $user_A$ requires $lib_1$ with 1.0 version, while $user_B$ requires $lib_1$ with 2.0 version, neither can lease the other's containers because of the version contradictions. 

    \item {\bf Create a vector to hold the libraries of each action}. For each lender action, the scheduler first filters the actions with common libraries with the lender action as the candidate actions. It then checks whether the libraries in the candidate actions are inconsistent with that in lender action (e.g., version contradiction). In that case, it will be removed from the candidate actions. Finally, the scheduler takes the union set of the libraries both in the lender action and rest candidate actions to form a libraries vector for distance calculation.

    \item {\bf Calculate the cosine distance between the lender action and the candidate actions as the similarity}. The filter logic is to select the top $n_L$ values of all the similarities and then take the corresponding actions as renters. If no candidate actions exist (for example, action-$NL$ is selected as lender action), $n_L$ action-$L$s without version contradiction will be added in random to be the renters. Besides, a number  $n_{NL}$ of action-$NL$s will also be selected in random as renters. 
\end{itemize}

Therefore, up to $n_L$ action-$L$s and $n_{NL}$ action-$NL$s will be selected as renters, and the inter-action container scheduler will wrap the renters' additional libraries into the image of the lender action. Meanwhile, the renters' code files will also be re-packed by $code\ encryption$ module for safety. $n_L$ and $n_{NL}$ are hyper-parameters and obviously their values affect the re-packing overhead and time. Their values should be set according to  (\ref{eq:6}), in which case all actions can get chances to be re-packed in lender containers.

\begin{equation}
\scriptsize
\begin{aligned}
\label{eq:6}
n_L = \min\{\frac{num(action\!-\!Ls)}{size(renter\ pool)}\},
n_{NL} = \min\{\frac{num(action\!-\!NLs)}{size(renter\ pool)}\}
\end{aligned}
\end{equation}

\begin{figure}
\centering
\centerline{\includegraphics[width=.96\columnwidth]{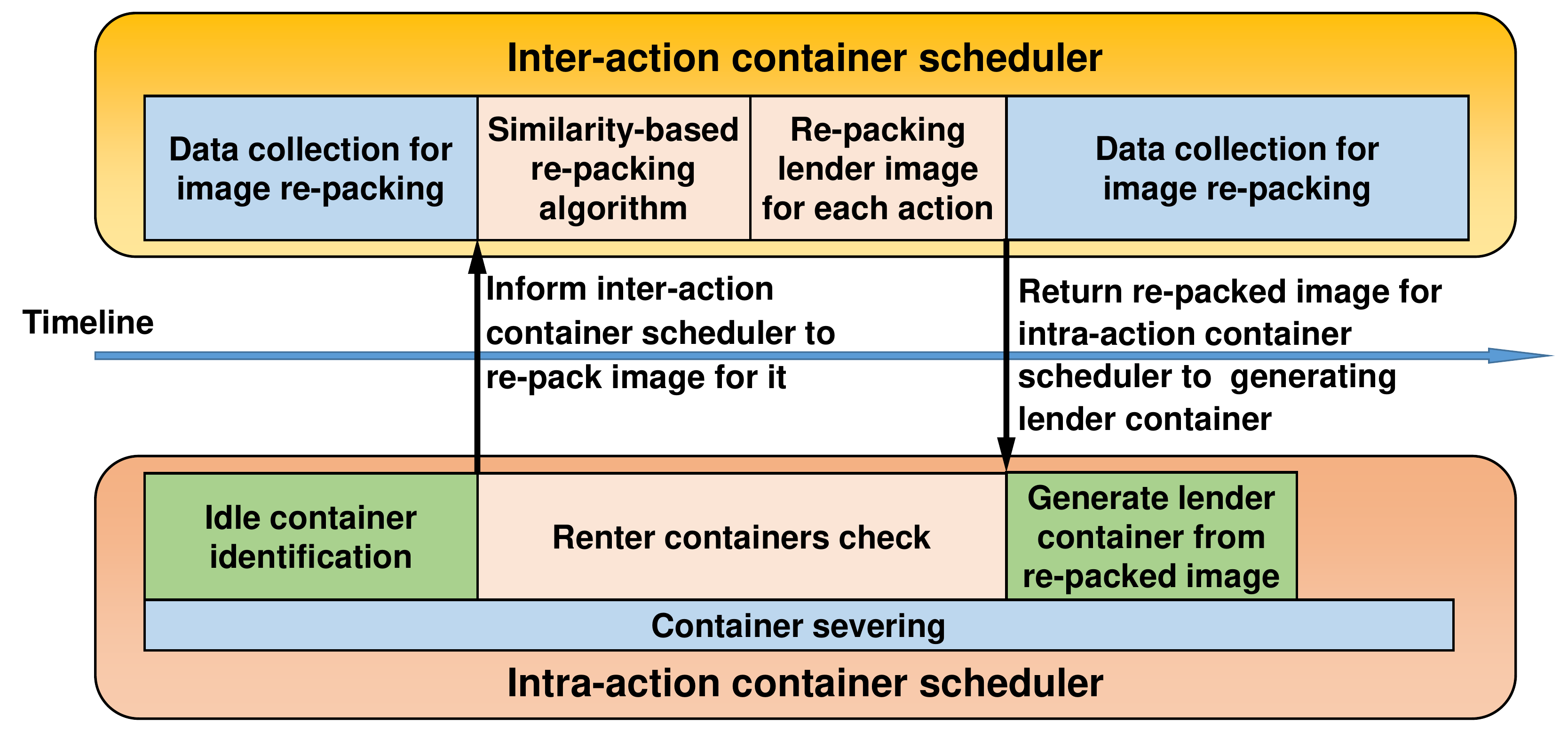}}
\caption{\label{fig:image}Timeline of Pagurus operations.}
\vspace{-4mm}
\end{figure}

Figure~\ref{fig:image} shows the re-packing operations along timeline. The inter-action container scheduler re-packs the lender containers images based on the data collection Image re-packing. The time cost in lender re-packing is hard to measure due to the uncertainty of the libraries vector cardinality.
With no doubt that higher cardinality, i.e., more libraries for re-packing, indicates longer re-packing time. 
But the re-packing phase for each lender action usually takes less than 10s for most actions according to our experiments. 
After re-packing, the images are committed to different intra-action container schedulers for creating lender containers. The overhead depends on the number of additional libraries to be installed. 
If some libraries take a relatively long time to re-pack, users will resort to submitting a virtual environment~\cite{openwhisk_pydoc} or custom container image~\cite{openwhisk_dockerdoc} to avoid the long installation time. In this case, Pagurus will adopt traditional CRIU to generate the containers, instead of re-packing.

The inter-action container scheduler deals with creating and updating of the re-packing image, while the intra-action container scheduler is responsible for managing the container pools, e.g., starting an executant container from default image, generating the lender container from the re-packed image. Unless the re-packed image is updated, the container only boots from it for the first time, and any subsequent container will use CRIU to accelerate the startup. 
The \textit{renter container check} module is designed to make sure that the runtime and libraries in the intra-action container scheduler are consistent with that in the inter-action container scheduler when performing the container re-packing.

\subsection{Security Guarantee}
In Pagurus, as a lender container may be shared by several renter actions, a natural and inevitable concern is on the security of  lender container. Meanwhile, as the code files of renter actions need to be re-packed in the shared container, the renters' security cannot be ignored.

For lenders' security guarantee, Pigurus explores the stateless nature of serverless computing to clean up user code and cache of the lender container before re-packing a lender image. No renter action can get any previous information about the action with the lender container.
For renters' security guarantee, Pigurus encrypts the renter action's code file by module $code\ encryption$ first before re-packing to prevent code disclosure.
The code encryption is divided into two parts. First, to protect the privacy of the users' files name, Pagurus  adopts a renaming strategy by renaming code files uniformly such as $\_\_main\_\_.py$, as adopted by OpenWhisk~\cite{openwhisk_doc_main}. Then, the environment folder for user actions will be encrypted into a ZIP file with the user password. Secondly, when a lender container is generated, all the renters' code files will coexist in this container. In this case, all the renters' folders need to be encrypted with inter-action container controller specified password to  protect the renters' privacy and code security in the lender container. It should be noticed that the cleanup and code decryption are executed in inter-action container scheduler, and therefore neither side can get any information about each other.


In conclusion, although Pagurus weakens the level of isolation, the code security and privacy required by isolation are still ensured and satisfied. Using encryption to secure data and files in cloud computing is quite common in practice~\cite{DBLP:conf/ccs/GoyalPSW06,DBLP:conf/iwqos/0001YLCXL13,DBLP:journals/ijnsec/TaiCH20}. So it is acceptable for actions to adopt encryption to address the security concern in container sharing.

\section{Inter-action Container Management}
\label{sec:manage}
In this section, we describe the steps of creating lender containers from idle containers, and using the borrowed container to run an action. 

\subsection{Generating a Lender Container}
\label{sec:generate}
\begin{figure}
\centerline{\includegraphics[width=.9\columnwidth]{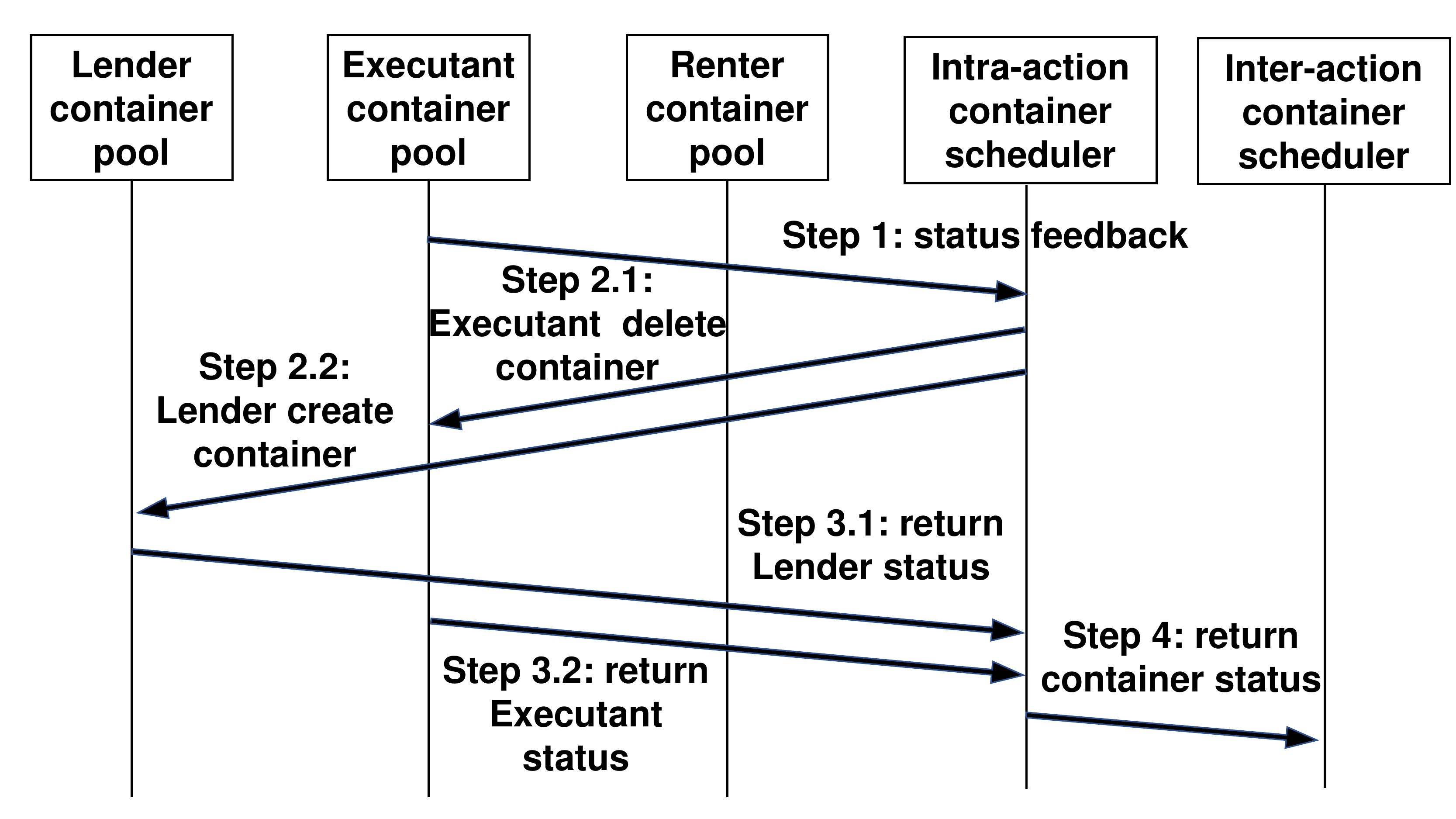}}
\caption{\label{fig:lender} Generating a lender container from an idle executant container.
}
\vspace{-1mm}
\end{figure}

If an executant container of an action is identified to be idle, its intra-action container scheduler will generate a {\it lender container} from the re-packed image returned by the inter-action container scheduler. Figure~\ref{fig:lender} shows the detailed workflow of generating a lender container.

As shown in Figure~\ref{fig:lender}, the executant containers of an action periodically feedback their status to the intra-action container scheduler. Based on the status of each container, the intra-action container scheduler identifies the redundant idle containers. Once an idle container is identified from the executant containers, the intra-action container scheduler re-packs the idle container to be a lender container (Step 2.1 and Step 2.2). In more detail, the idle container is deleted from the executant container pool, and the corresponding lender container is added to the lender container pool. This information is then feedbacked to the intra-action container scheduler (Step 3.1 and 3.2), so that the scheduler is aware of the change.
In the last step, the intra-action container scheduler informs the inter-action container scheduler of the change (Step 4). In this way, other actions are able to borrow the container through the inter-action container scheduler.




\subsection{Renting a Container from Other Actions}
\label{sec:renting}
\begin{figure}
\centerline{\includegraphics[width=.9\columnwidth]{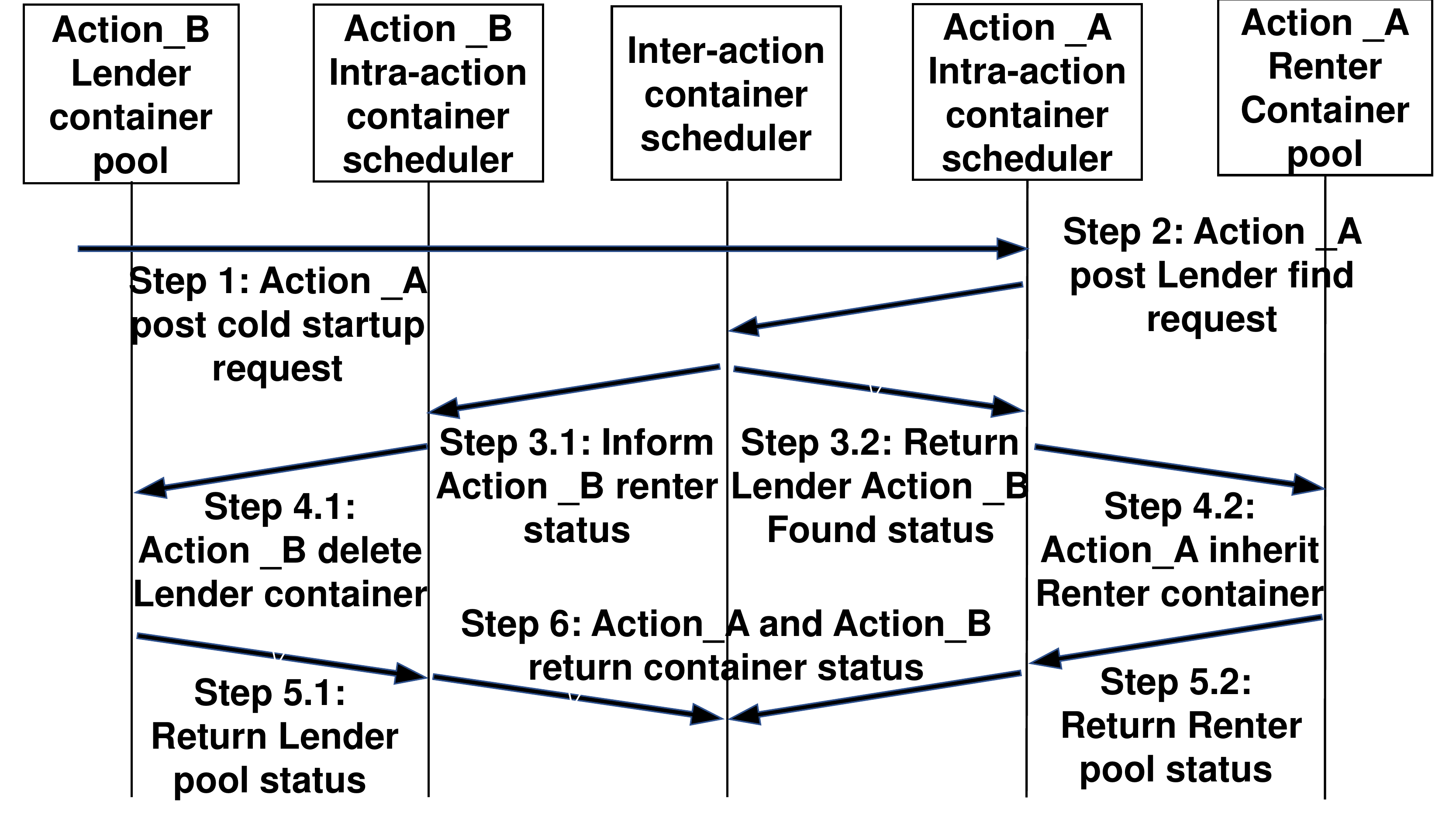}}
\caption{\label{fig:renter} The steps of Action\_B rents a container from Action\_A. 
}
\vspace{-3mm}
\end{figure}

When an action {\it ACT} needs a container to run but there is no free warm container for it, 
its intra-action container scheduler submits a rent request to the inter-action container scheduler. 
If there exists such lender container that is already prepared for {\it ACT} by other actions, 
the container is changed to be a {\it renter container} of {\it ACT}, and is put in the {\it renter container pool} of {\it ACT}.
Figure~\ref{fig:renter} shows the detailed steps of renting a container from other actions.

Among these steps, it is crucial to guarantee the lender's container delivery and ensure the information safety of the lender. In Step 3, the inter-action container scheduler deletes the code and data of $action_B$, as well as the other renters' code file in the lender container, and deciphers the code file of $action_A$. Then inter-action container scheduler will inform $action_B$ and $action_A$ to prepare for container transferring (Step 3.1 and Step 3.2).
Once $action_A$'s intra-action container scheduler receiving return container status, it will schedule this lender container to its renter container pool (Step 4.2) and $action_B$'s lender container pool will clear related status and information (Step 4.1). Meanwhile, the management privilege of the lender container is transferred from $action_B$'s intra-action container scheduler to $action_A$'s intra-action container scheduler. 

The code cleaning of $action_B$ and the code decryption of $action_A$ are executed in parallel. While the overhead of cleaning is less than the time cost of code decryption, the overhead of code cleaning is hidden from users.

\begin{figure}
\centerline{\includegraphics[width=.86\columnwidth]{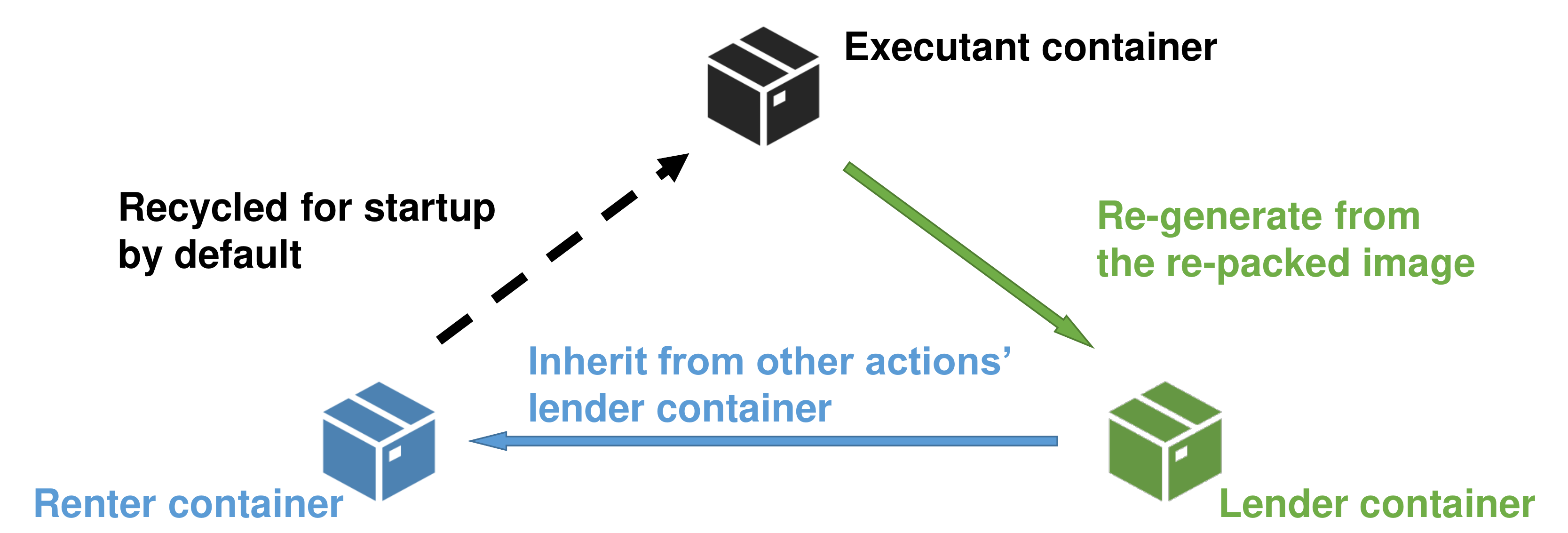}}
\caption{\label{fig:transfer} The state transition diagram of three containers. 
}
\vspace{-3mm}
\end{figure}

Based on Section~\ref{sec:generate} and \ref{sec:renting}, we can summarize the state transitions for three different containers in Figure~\ref{fig:transfer}. The action is executed in the executant container by default, and all cold startup containers are managed in the executant container pool. Lender container is transformed from an idle executant container identified by its intra-action container scheduler to a shared container re-generated from the re-packed image. Renter container inherits from other actions' lender container to make queries get executed without container cold startup. All these containers will keep running until timeout to be recycle.

\subsection{Recycling Containers in Different Pools}
In serverless computing, when the load of an action drops, some warm containers for the action are recycled to save resources. Recycling is done by monitoring the status of the containers. If a container does not receive new requests for a time period (60s in OpenWhisk), the container is recycled. The recycling is only done by 
setting a timeout period for each container. If a container does not operate in the timeout period, the container will be recycled. 
This recycling policy is not able to be used in Pagurus directly as there are three types of containers in Pagurus.

\begin{figure}
\centerline{\includegraphics[width=\columnwidth]{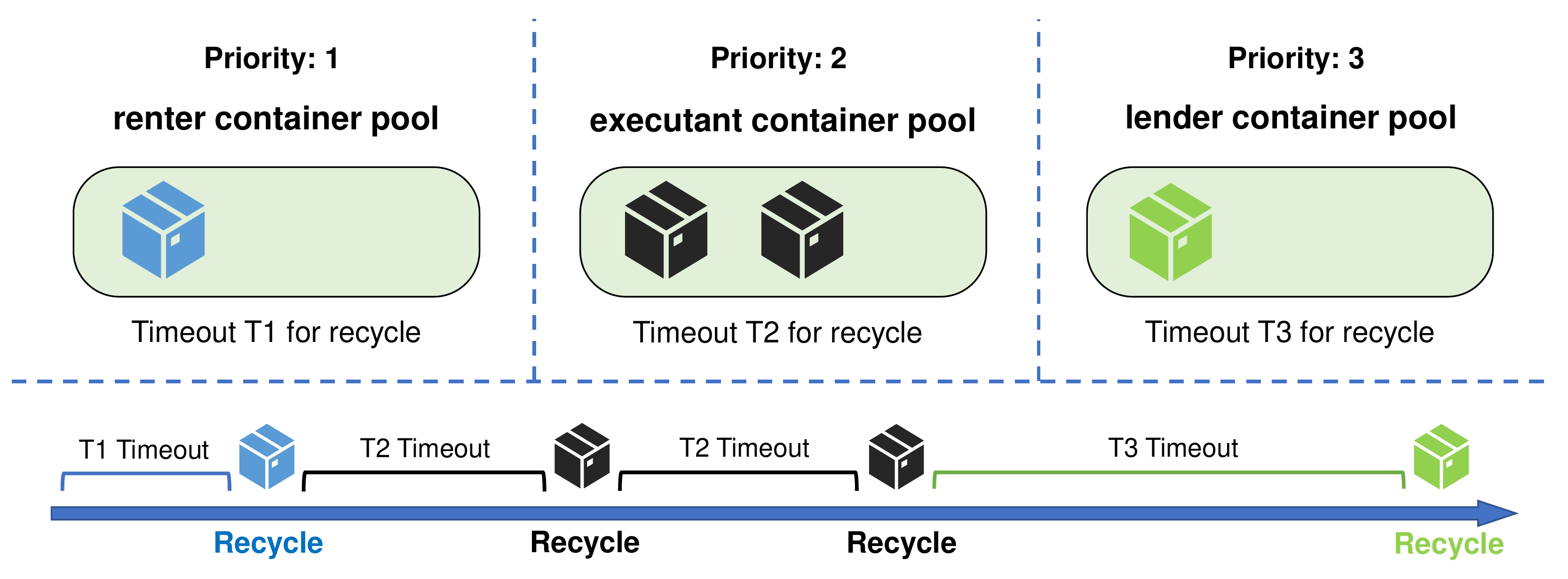}}
\caption{\label{fig:recycle}Recycling thee three types of containers.
}
\end{figure}

Therefore, we design a priority-based recycle strategy for Pagurus.
In this strategy, the inter-action container scheduler manages the recycling of all its containers, including executant containers, lender containers, and renter containers. 
For an action, Pagurus recycles the renter containers before all the other containers, and recycles lender containers after all the containers.
The design philosophy here is that an action does not need extra rent containers when its containers tend to be recycled. Figure~\ref{fig:recycle} shows the order of recycling containers when the load of an action drops.

Specifically, we set different timeout periods for the three types of the containers. The renter container pool has the minimum timeout period ($T1$ in Fig.~\ref{fig:recycle}). The executant container does not store information and libraries for other actions, the container recycle does not affect the scheduling of the intra-action container scheduler. 
The timeout period of executant containers ($T2$ in Fig.~\ref{fig:recycle}) is slightly larger than the timeout period of the lender containers. 
Because the lender containers re-pack additional libraries for multiple actions, even if executant and renter containers are all recycled, it can also meet the invocation of the action. For the above reasons, the lender containers have the maximum timeout period ($T3$ in Fig.~\ref{fig:recycle}). 

In our current implementation, we set the timeout periods for the renter containers, executant containers, and lender containers to be 40s, 60s, and 120s by default. 



\section{Evaluation of Pagurus}
\label{sec:eval}
In this section, we first evaluate the performance of Pagurus in reducing the end-to-end latencies of applications when there is no warm containers for them. Then, we discuss the 
possibility of Pagurus in eliminating the cold startup, the effect of Pagurus in supporting bursty load, and its effect in integrating with the orthogonal techniques. 


\subsection{Experimental Setup}
In the experiment, we evaluate Pagurus based on a 2-node cluster described in Section~\ref{sec:setup}. In the 2-node experimental cluster, they serve for inter-action container scheduler re-packing. Although we only use a small scale cluster in this section,  Section~\ref{sec:multi-nodes} reveals the situation in large scale Clouds while serverless computing platforms often manage the containers on each node independently. While containers are in the process context, containers are not supposed to be migrated to other nodes in most cases.


\begin{figure}
\centerline{\includegraphics[width=.8\columnwidth]{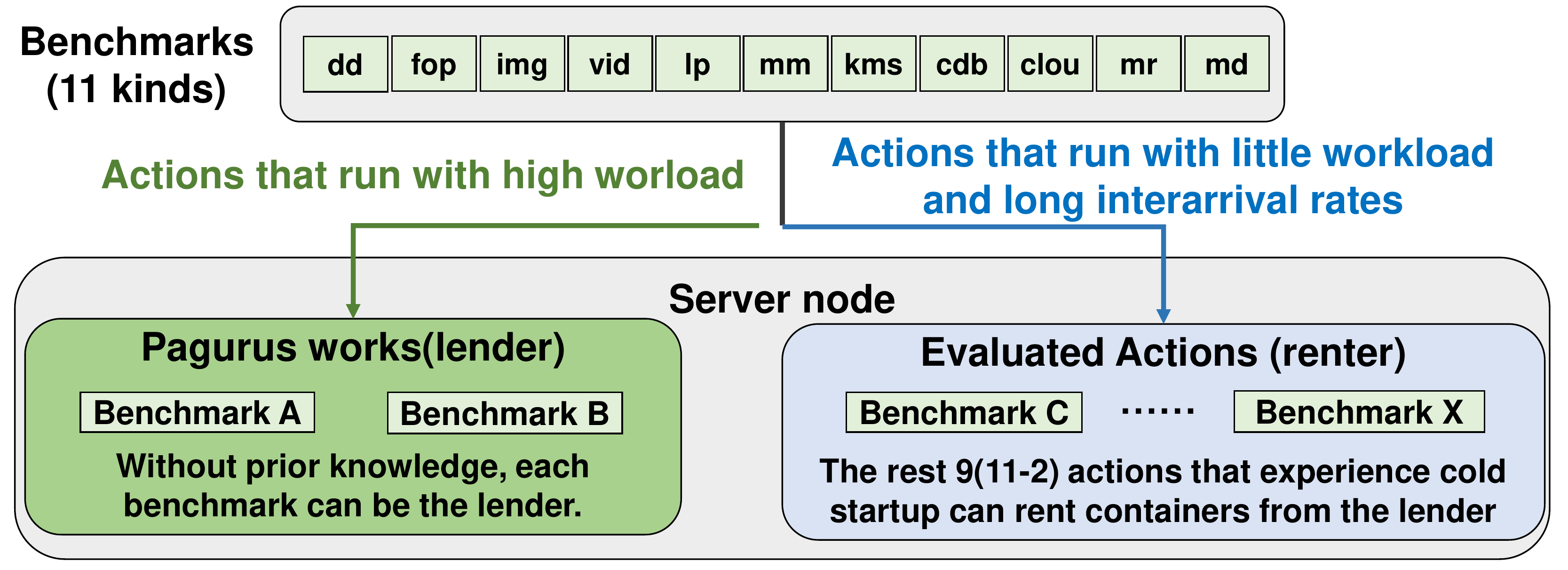}}
\caption{\label{fig:background}The configurations of actions running in the background.
}
\end{figure}

We use representative serverless computing benchmark suites FunctionBench~\cite{functionbench} and Faas-profiler~\cite{DBLP:conf/micro/ShahradBW19} to evaluate Pagurus. Table~\ref{tab:benchmarks} lists the used benchmark workloads.
In the following experiment, we set the maximum number of containers in the renter pool to be 2 and randomly run two benchmarks in the background with high loads to simulate the real-system situation. To better understand the background configuration, we make a schematic diagram, as shown in Figure~\ref{fig:background}. In a real system, there are some long-running services and some occasional queries on the same serverless computing platform, and the services running in the background are also uncertain. So there are total $C_{11}^2 = 55$ combinations of experimental configurations for Pagurus when we randomly select two benchmarks to be the lender.
In each experimental configuration, we run each benchmark for 100 times  by invoking the benchmark once every 60 seconds. In this way, the benchmark suffers from the cold container startup in all the 100 tests.
In the following experiment, we collect the end-to-end latencies of the 100 tests for each benchmark based on the above experimental setup.



\subsection{Reducing End-to-end Latency}
We show the effectiveness of Pagurus in reducing the end-to-end latency of a benchmark in this subsection.
In this experiment, for each benchmark, we randomly select two of the other 10 benchmarks to be the lenders in the background for Pagurus. We compare Pagurus with Apache OpenWhisk~\cite{openwhisk} and the restore-based method~\cite{CRIU}. OpenWhisk creates a new container for a benchmark from the corresponding container image and startups the new container. Restore-based method stores the checkpoint of the container in the memory, and restores the checkpoint from the main memory when needed.


\begin{figure}
\centerline{\includegraphics[width=.9\columnwidth]{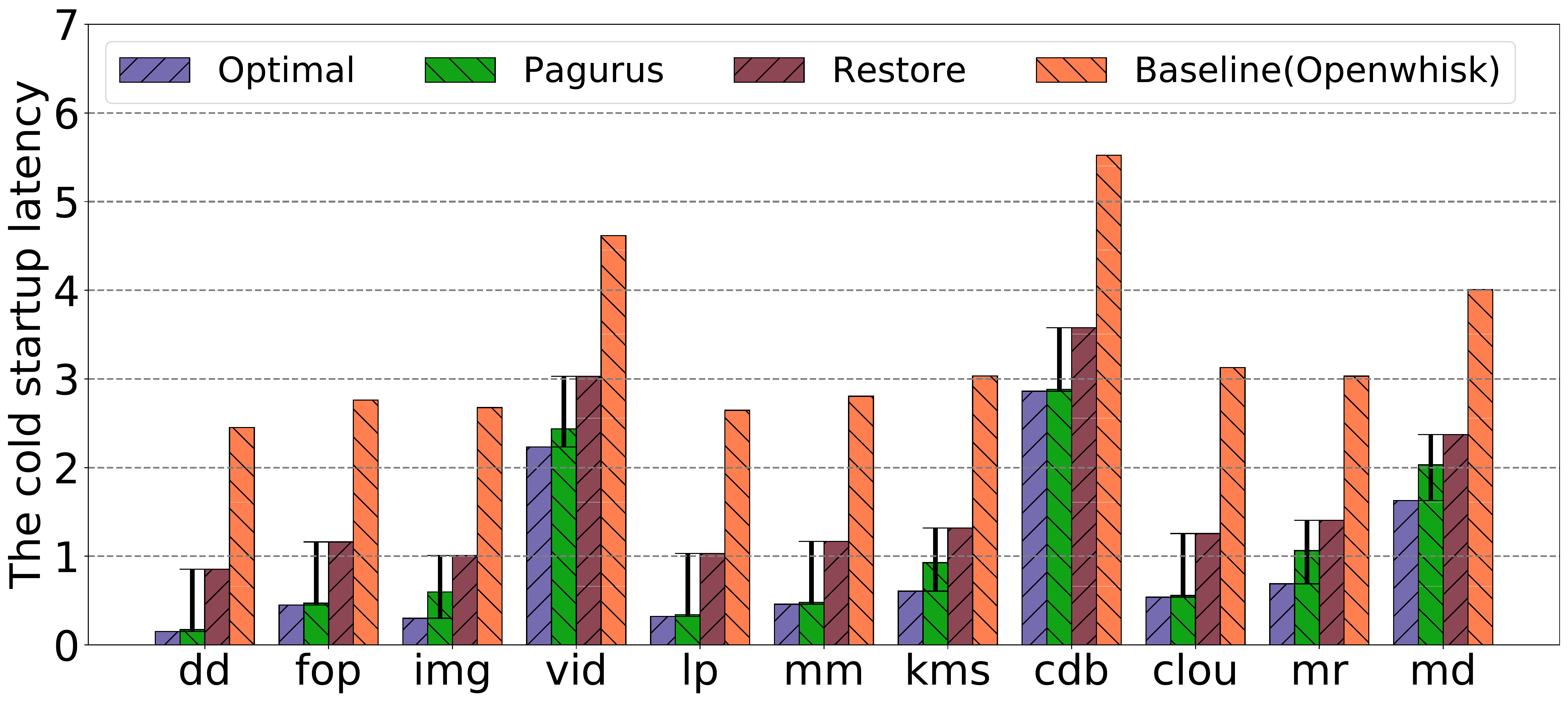}}
\caption{\label{fig:coldcompare}The end-to-end latencies of the benchmarks when they suffer from cold startup with OpenWhisk, Restore-based method, and Pagurus.
}
\vspace{-3mm}
\end{figure}

Figure~\ref{fig:coldcompare} shows the end-to-end latencies of the benchmarks with OpenWhisk, Restore-based method, and Pagurus. 
In the figure, the ``optimal'' reports the latencies of the benchmarks when they get warm containers directly. As shown in the figure,
all the benchmarks achieve the shortest end-to-end latencies with Pagurus compared with OpenWhisk and Restore-based method.
In the best case where actions get lender containers, Pagurus reduces the end-to-end latencies of the benchmarks by 75.6\% and 51.9\% compared with OpenWhisk and the restore-based method respectively. When compared to the optimal scenario, Pagurus only introduces 0.48\% longer end-to-end latency on average.


Pagurus greatly reduces the end-to-end latencies because it schedules the idle shared containers to speed up the actions that may suffer from cold container startups. If an action query is hosted in a shared container, the container startup phase for the query is skipped and only the user-specific code initialization is needed. 
According to our measurement, Pagurus schedules a lender container to a query in less than 15\textmu s, and completes the container cleaning and application-specific code initialization in less than 10ms.

{\hcolor
The restore-based method is also able to reduce the end-to-end latencies of the benchmarks compared with OpenWhisk. This is mainly because it eliminates the overhead of creating the new container images. However, it consumes large memory space and still results in longer end-to-end latencies of the benchmarks compared with Pagurus.
}

\subsection{Eliminating the Container Cold Startup}
It is possible that there is not a renter container for an action.
For an action, the extra libraries  (software  libraries) it packs determine the probability that it will skip a cold startup.
The probability of eliminating the cold startup is an important indicator that reflects the effectiveness of Pagurus. 

\begin{figure}
\centerline{\includegraphics[width=.86\columnwidth]{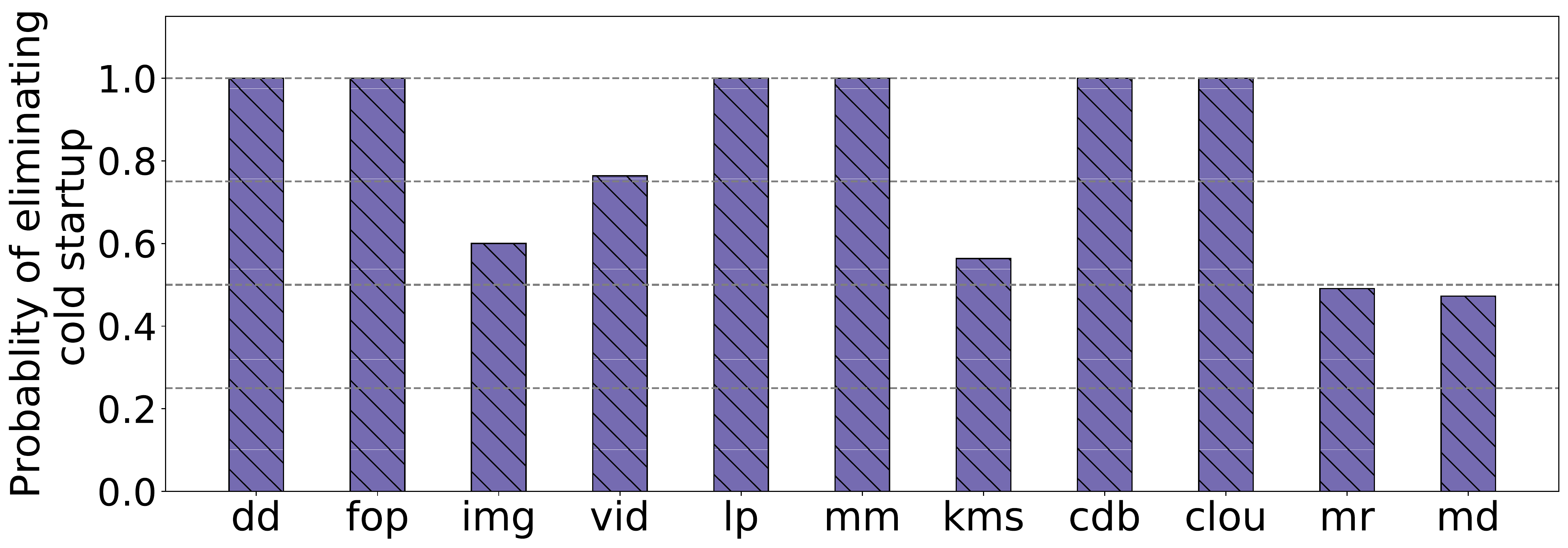}}
\caption{\label{fig:recoldfre}The probability of eliminating cold startup with Pagurus.
}
\end{figure}

In this experiment, for each benchmark, we run it in $C_{10}^2 = 45$ experimental setups. For each setup, we select 2 out of the 10 benchmarks as the renters. 
Figure~\ref{fig:recoldfre} shows the percentages of the $C_{10}^2 = 45$ setups in which the benchmarks skip the cold container startup.

Observe from Figure~\ref{fig:recoldfre}, Pagurus eliminates all the cold container startup for {\it dd}, {\it fop}, {\it lp}, {\it mm}, {\it cdb} and {\it clou}, 
because these benchmarks can always rent containers from the lenders. 
They can always find the lenders because these benchmarks do not require additional libraries to initialize the software environment. In this case, the container re-packing algorithm is able to pack the redundant idle containers of any actions to be its renter containers.

For the benchmarks that require extra libraries ({\it img}, {\it vid}, {\it kms}, {\it mr}, and {\it md}),
the possibility of eliminating the cold startup depends on the libraries similarity of the lender actions and renter actions. 
The more common and popular the additional libraries required in the action, the higher the probability that it will be re-packed by the lender actions. For instance, in 77.3\%, 59.1\% and 57.6\% of the configurations, Pagurus eliminates the cold container startup for {\it vid}, {\it kms}, and {\it img} respectively. This is because these benchmarks mainly use the shared {\it Pillow} and {\it sk-learn} software packages. However, for {\it mr} and {\it md}, due to the unpopular of packages they used, lender actions take lower priority to pack it in lender containers. The decision leads to the relatively low probabilities (34.8\% and 36.4\%) of eliminating cold startup for  {\it mr} and {\it md}. 

\begin{figure}
\centerline{\includegraphics[width=.6\columnwidth]{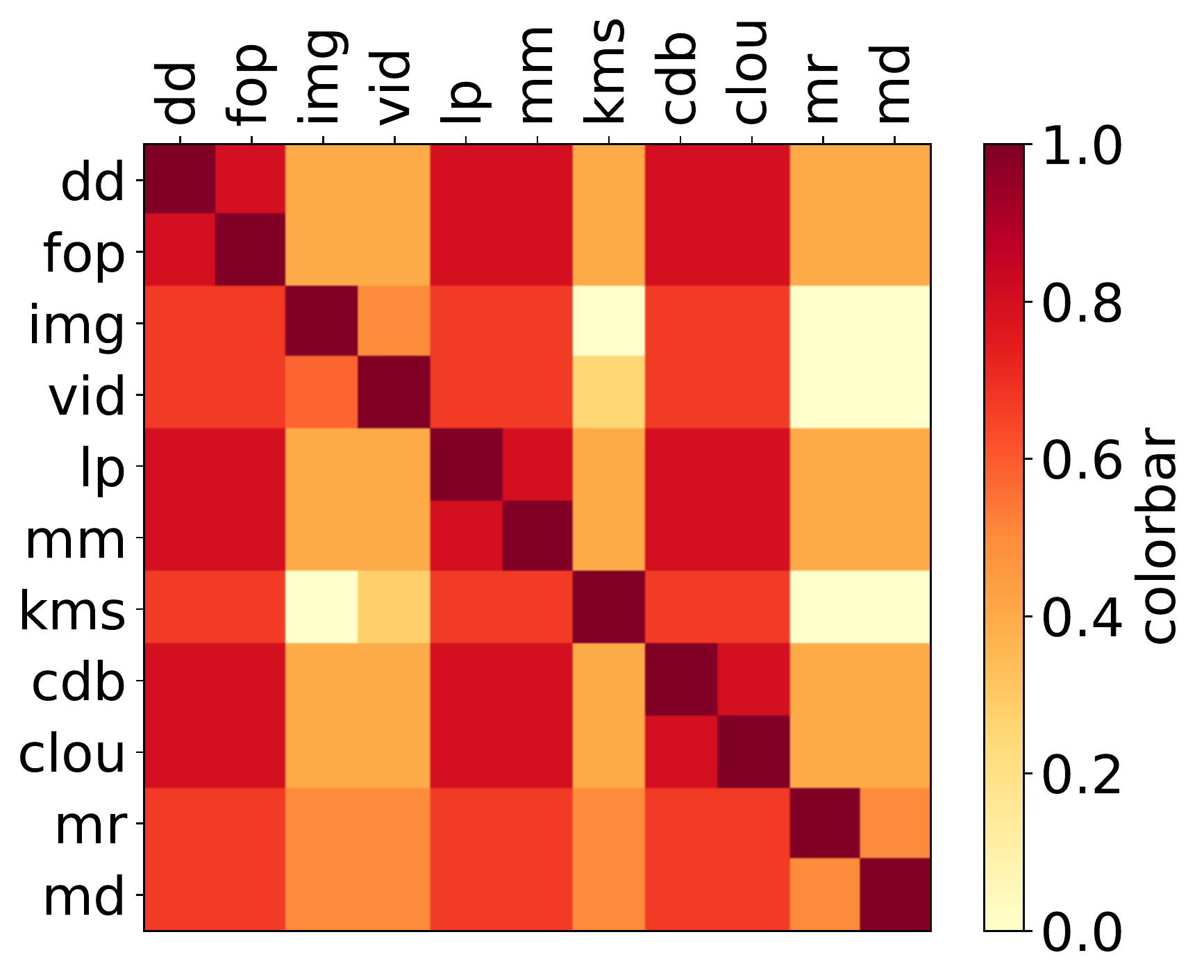}}
\caption{\label{fig:heat}The benchmark similarities in the container re-packing algorithm.}
\vspace{-3mm}
\end{figure}

To better understand this problem, Figure~\ref{fig:heat} shows the heat map of the benchmark similarities in the container re-packing algorithm. In the figure, the very small square in row {\it vid} and column {\it img} represents the possibility that {\it vid} serves as the lender for {\it img} rental. The small square in the row {\it img} and column {\it vid} represents the possibility that {\it vid} serves as the renter for {\it img}. 
{\hcolor Observed from the figure, all the benchmarks do not tend to re-pack containers for {\it mr} and {\it md}. These results explain the reason that Pagurus shows a relatively low possibility in eliminating cold container startup for {\it mr} and {\it md}. An alternative method to further resolve this problem is taking the prior knowledge into consideration. Another way is increasing the number of renters that each lender can choose.}


{\hcolor It should be noticed that the heat map in Figure~\ref{fig:heat} is asymmetric, because the benchmarks rely on different software packages. Assume an action $ACT_1$ relies on software packages $lib_1$, $lib_2$ and another action $ACT_2$ relies on software packages $lib_1$. In this case, the containers of $ACT_1$ have all the packages for $ACT_2$. At the same time, the containers of $ACT_2$ only have half of the software packages for $ACT_1$. The possibilities of re-packing the containers of $ACT_1$ for $ACT_2$, and re-packing the containers of $ACT_2$ for $ACT_1$ are different.} 


Renter containers can be used by multiple actions and not all the benchmarks can always skip the code container startup. 

\subsection{Integrating with Work on Reducing Cold Startup Time}
\begin{figure}
\centerline{\includegraphics[width=.92\columnwidth]{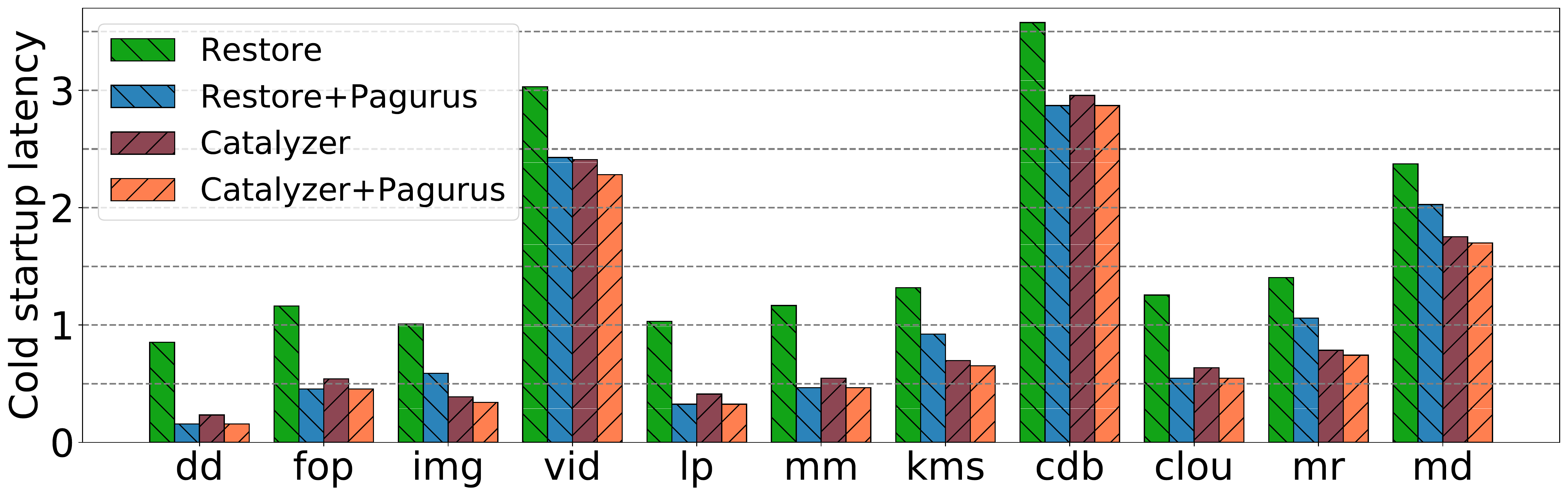}}
\caption{\label{fig:combine}The average cold startup latency when Pagurus is integrated with Restore-based method, and Catalyzer respectively.}
\vspace{-1mm}
\end{figure}

While Pagurus eliminates the container cold startup, it can be integrated with prior work proposed to reduce the container cold startup time. In this subsection, we integrate Pagurus with Restore-based method and Catalyzer~\cite{du2020catalyzer} respectively, and report their performance. For each benchmark, we still run it in the 45 experimental setups. In each experimental setup, we launch the benchmark 100 times with an interval of 60 seconds. Figure~\ref{fig:combine} shows the average container startup time for each benchmark of the $45\times 100 = 4,500$ tests.

Observed from Figure~\ref{fig:combine}, {\it Restore+Pagurus} reduces the average container startup time of the benchmarks by 43.4\% on average compared with the original Restore-based method; {\it Catalyzer+Pagurus} reduces the average container startup time by 12.2\% on average compared with Catalyzer. 
Pagurus is able to further reduce the average container startup time, because it is able to totally skip the container startup phase for the benchmarks sometimes.
Even if no appropriate lender container returns, Pagurus will not slow down the container startup.
Therefore, Pagurus can be integrated with prior work to further reduce the average cold startup time.




\begin{figure}
\centering
\subfigure[mm]{
\includegraphics[width=.22\textwidth]{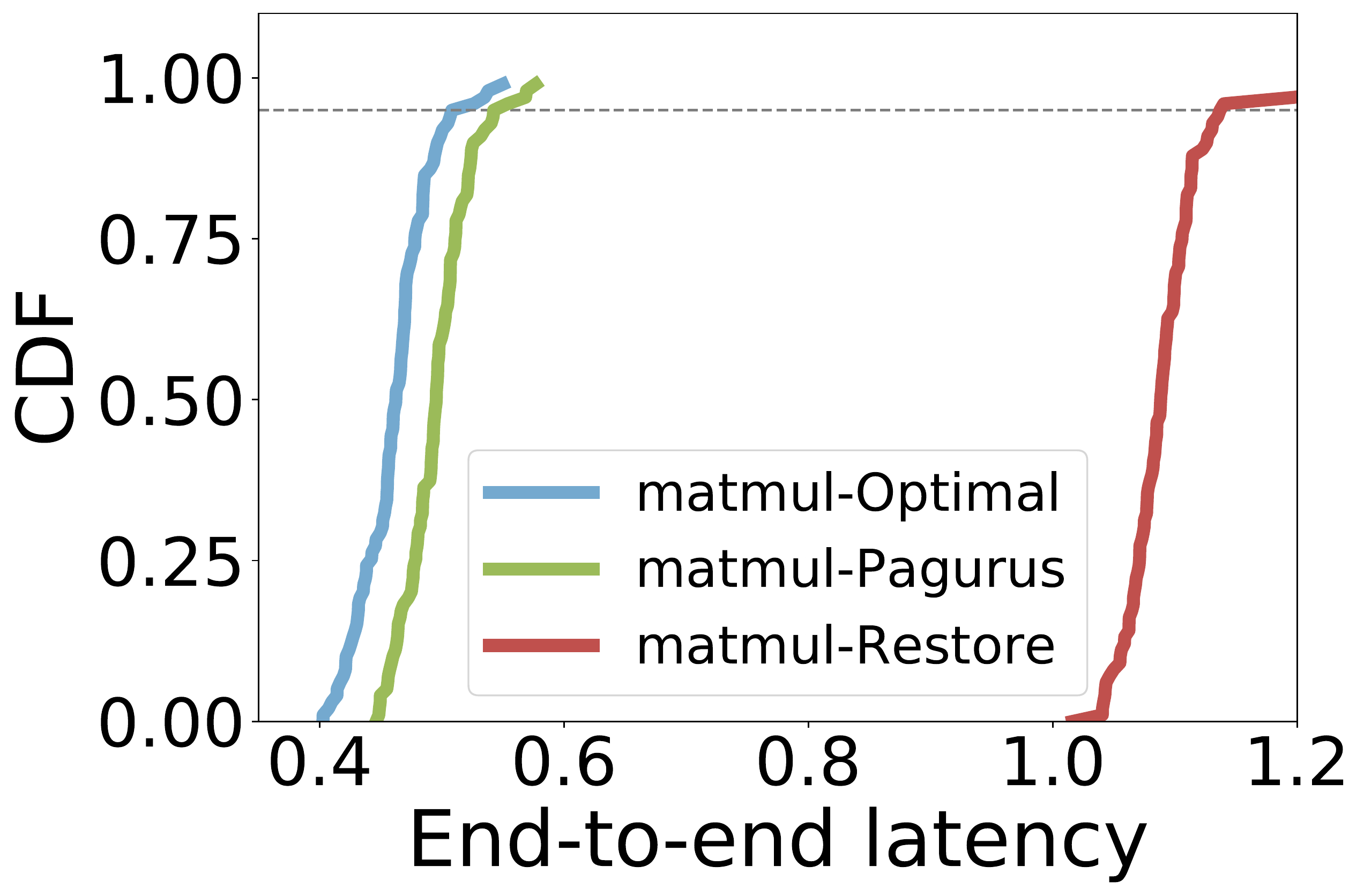}
}
\subfigure[img]{
\includegraphics[width=.22\textwidth]{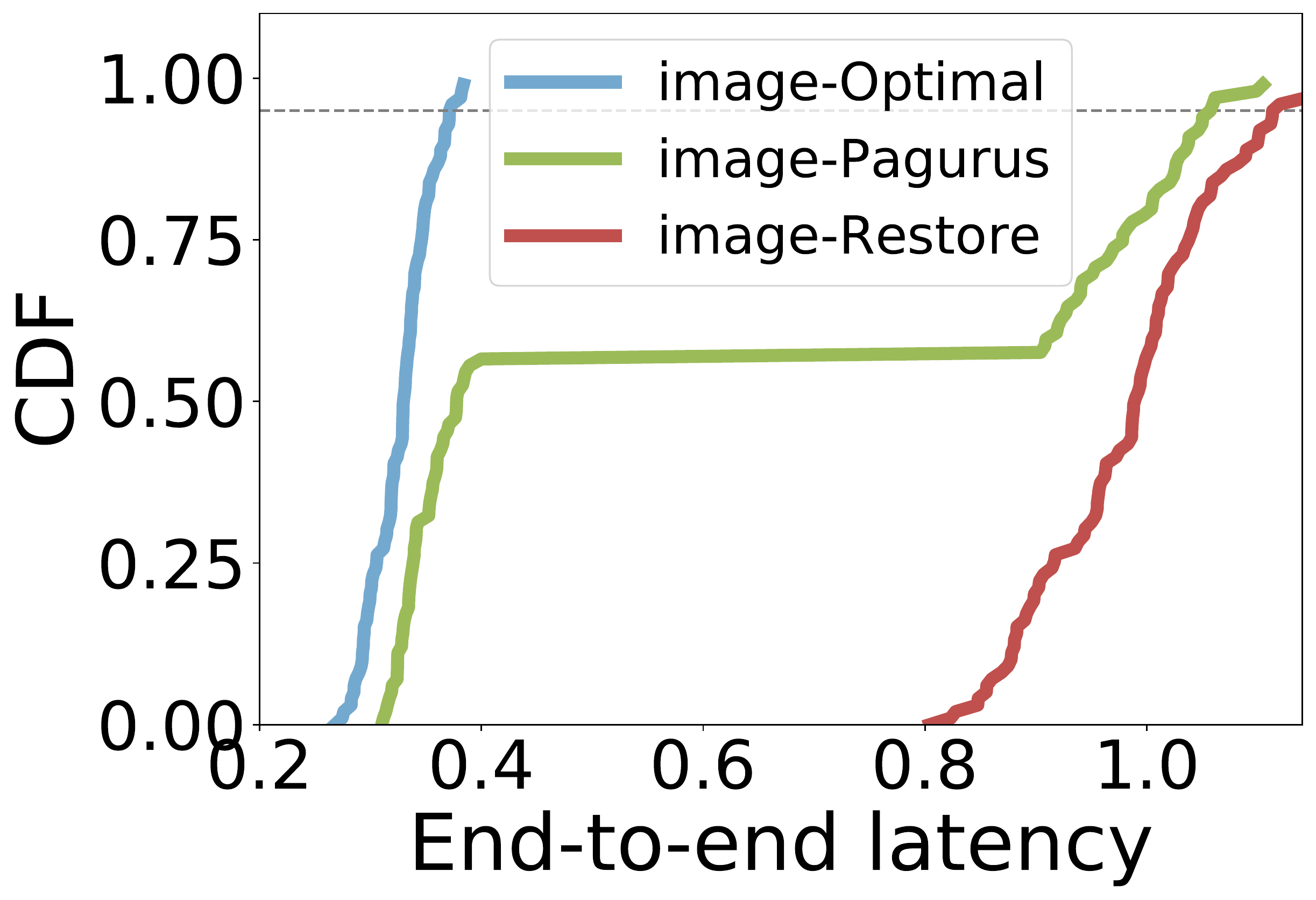}
}
\caption{The cumulative distribution of the benchmarks' cold startup end-to-end latencies in Pagurus and Restore.}
\label{fig:cdf}
\vspace{-3mm}
\end{figure}

Figure~\ref{fig:cdf} shows the cumulative distribution of the end-to-end latencies of {\it mm} and {\it img} with {\it restore-based method} and {\it Restore+Pagurus}. 
In the figure, ``optimal'' shows the latencies of the benchmarks when all the containers are warm. 
By integrating Pagurus with the restore-based method, the end-to-end latencies of all the 4,500 tests of {\it mm} are greatly reduced. Meanwhile,  52.1\% of the queries of {\it img} show much shorter end-to-end latencies.
In {\it mm}, action queries completely skip the container startup, thus overhead of cold startup is eliminated. 
While for {\it img}, about 52.1\% of the action queries can eliminate cold startup overhead, the rest still have to experience the container startup phase with the restore-based method.

For Pagurus, we can observe a large discontinuity in Figure~\ref{fig:cdf}(b) as Pagurus helps most of the queries to skip the container startup phase to reduce the latency, and only a few of them still suffer from cold startup problem.

\begin{figure}
\centerline{\includegraphics[width=.84\columnwidth]{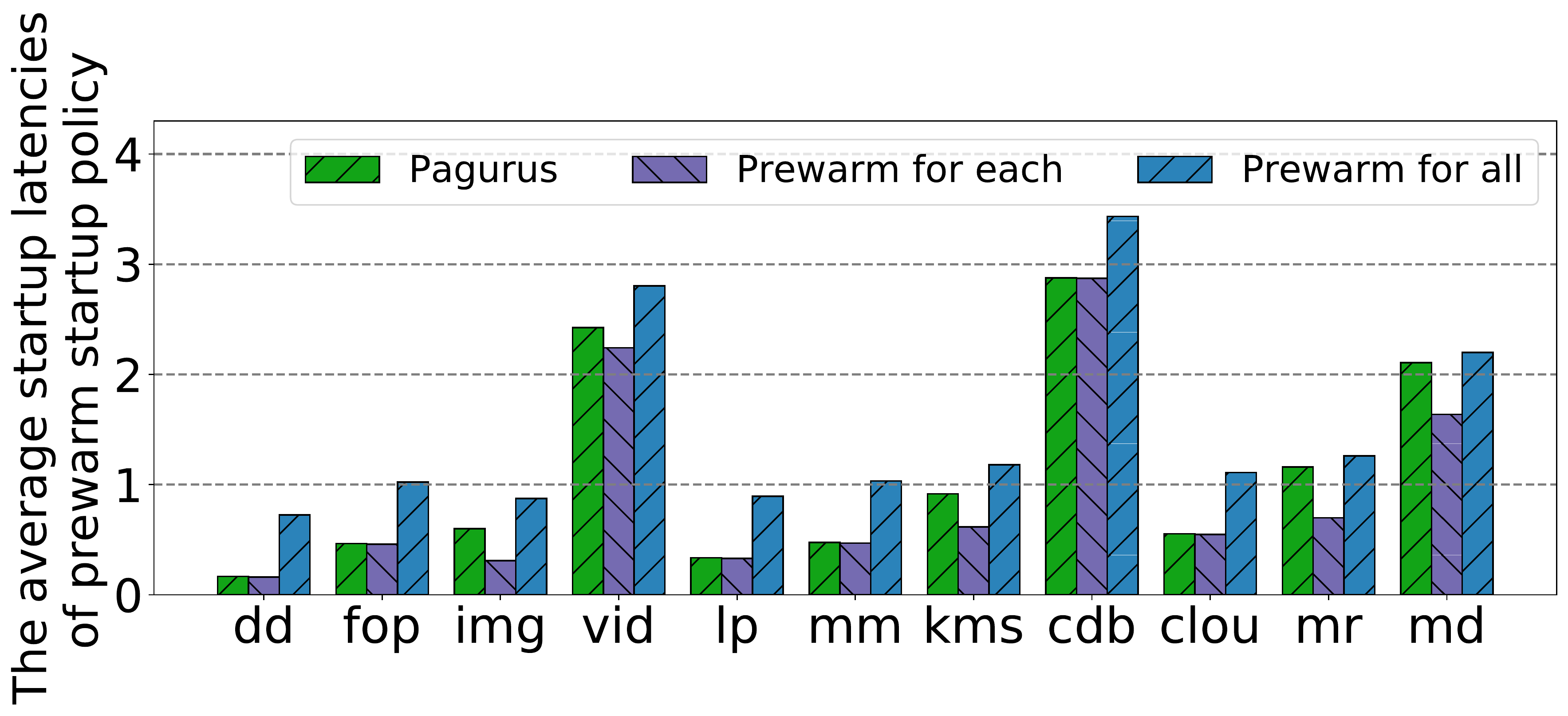}}
\caption{\label{fig:prewarm_compare} The average cold startup latencies of the benchmarks with prewarm startup policy. 'Prewarm for each' means that each action can get one prewarmed container, and 'prewarm for all' means all actions can initialize one specific container created from a common cache.}
\vspace{-3mm}
\end{figure}

Traditionally, we can also integrate the prewarm startup (introduced in Section~\ref{sec:related}) policy with these advanced container startup techniques. From the Figure~\ref{fig:prewarm_compare}, we can observe that Pagurus still performs better than `prewarm for all' method. It is because that the libraries in the specific prewarmed container may conflict with the user action. In this case, these specific prewarmed containers cannot be initialized, making user actions experience the cold startup. Even though the `prewarm for each' method shows less end-to-end latency than Pagurus due to the prewarmed containers continuously running in the background, additional 2.75GB memory resources are required by it. So, although with comparatively high performance, `prewarm for each' method is unpractical because of the extremely high resource usage.

\subsection{Supporting Bursty Loads}
Traditional serverless platforms (e.g., OpenWhisk) fail to support bursty workload without causing QoS violation due to the long latency during the cold container startup. 
Pagurus is able to support the smooth process of the bursty workload of an action through the inter-action container sharing.
\begin{figure}
\centerline{\includegraphics[width=.8\columnwidth]{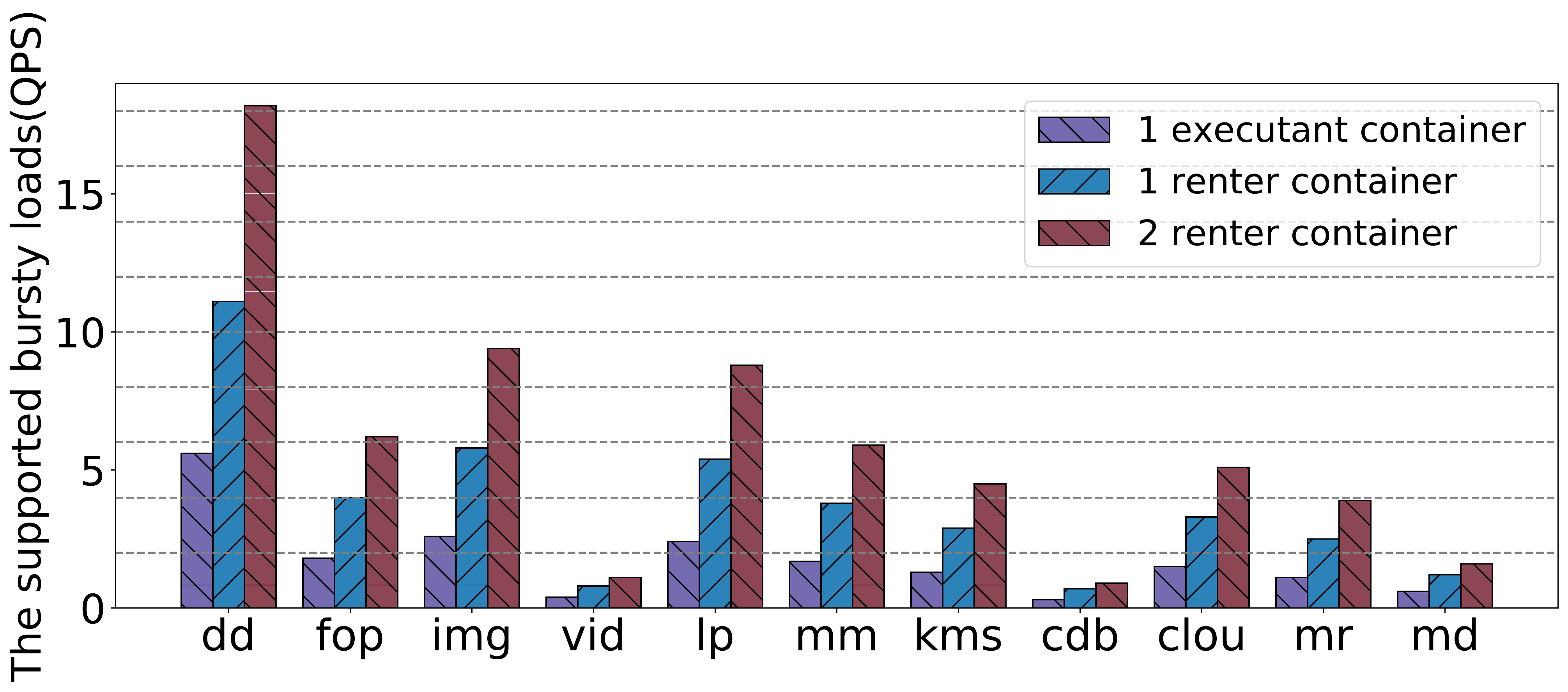}}
\caption{\label{fig:burst} The supported bursty loads of the benchmarks with Pagurus.}
\end{figure}

Figure~\ref{fig:burst} shows the supported bursty loads of the benchmarks without causing the QoS violation, when Pagurus allows the benchmark to rent 1 or 2 more containers. 
Observed from Figure~\ref{fig:burst}, for all the benchmarks, Pagurus is able to support 3$\times$ of the bursty loads if the benchmarks are able to rent 2 more renter containers from other actions. This is mainly because the overhead of renting containers from other actions is much lower than creating new containers.
\begin{figure}
\centerline{\includegraphics[width=.8\columnwidth]{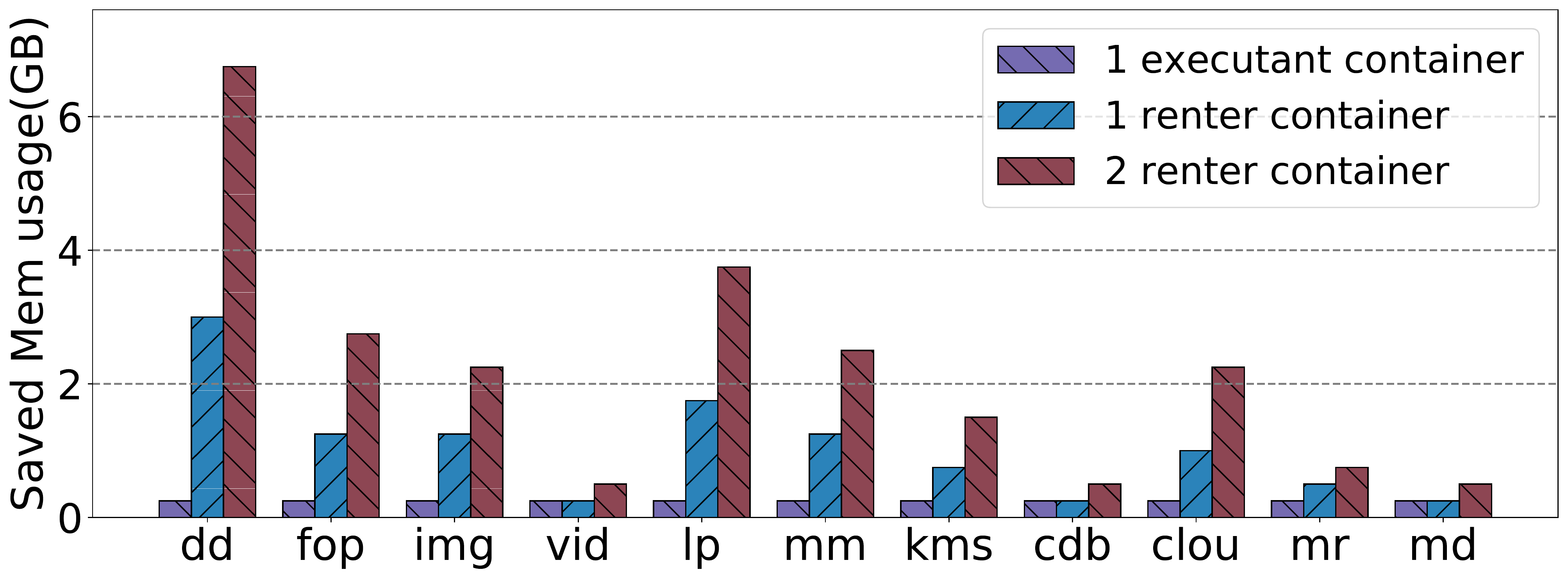}}
\caption{\label{fig:resource} The size of the reduced memory usage to support the bursty loads with Pagurus compared with OpenWhisk.}
\vspace{-3mm}
\end{figure}

Besides, Pagurus also reduces the consumed memory space to support the bursty loads of the benchmarks compared to OpenWhisk. 
To suppose the bursty load with OpenWhisk, a straightforward method is maintaining more warm containers.
However, these containers consume large main memory space. 
On the contrary, with Pagurus, there is no need to launch additional containers to support the bursty loads. 
Figure~\ref{fig:resource} shows the size of the main memory space saved with Pagurus. As shown in the figure, 
Observed from the figure, in Pagurus, 0.25GB to 3GB of the memory is saved in the case of 1 renter container, 
and 0.5GB to 6.75GB of the memory in the case of 2 renter containers compared with OpenWhisk.

\subsection{Overheads in Pagurus}
\begin{table}
  \centering{}
  \caption{\label{tab:overhead} Time and space overheads introduced by Pagurus}
  \scriptsize
  \begin{tabular}{c|c|c}
    \toprule
     Type & Position & Time/Resource Overhead\\
    \hline
     Encrypted code files size & Running lender container & 4.3125KB \\
     Re-packed image size & Creating lender container & 485MB \\
     Re-packing image time & Creating lender container & 6.647s\\
     Checkpoint files size & Creating container & 332KB \\
     CPU overhead & Re-packing container & 1.61\% \\
    \bottomrule
  \end{tabular}
\end{table}
Table~\ref{tab:overhead} shows the overheads introduced by Pagurus. 
As shown in the table, Pagurus incurs five types of overheads.
In Pagurus, the lender container stores the renters' encrypted code files as an extra operation, and decrypt the corresponding renter's code file when eliminating cold startups. This approach only takes 4.3125KB of space to save information and less than 10ms to decrypt, which is far less than about 200ms of database transmitting. 
The operation of packing images is introduced in the generation of lender containers. The extra packed images experience creating by average 6.647s, and 485MB of space is allocated for storing. {\hcolor The preparation of lender containers is done asynchronously and does not result in the long end-to-end latency.} Containers only need to boot from images for the first time. In other cases, containers get accelerated in startup by checkpoint files which takes average 332KB of space to store. 
Besides, the re-packed images and the checkpoint files will be recycled when the corresponding actions are not invoked.

The most important part of the overhead is the CPU usage when Pagurus re-packing containers. The experiment shows that when the inter-action container scheduler re-packs the container image for an action, the average CPU utilization on the node is only about 1.61\%. After taking the communication and synchronization between nodes into account, the re-packing phase will consume about 2.4\% of the CPU resource. If we limit the CPU usage of re-packing to less than 10\% of the server, each node can re-pack container images for about 34 actions during the data collection in 1 minute.
To conclude, the overhead incurred by Pagurus is negligible.

\section{Conclusion}
This paper presents Pagurus,a container management system for serverless to eliminate container cold startup by inter-action container sharing. We implement the design by introducing three unique container pools, {\it lender containers}, {\it executant containers} and {\it renter containers}. The inter-action container scheduler cooperated with intra-action container schedulers in each action, enable containers scheduled between different actions to reduce container cold startup. The evaluation result shows that Pagurus can significantly eliminate the cold startup.
Besides, Pagurus can also be integrated with several container technologies to minimize the container startup overhead of serverless computing.

\bibliographystyle{plain}
\bibliography{reference}

\begin{thebibliography}{10}

\bibitem{lambda}
\url{aws.amazon.com/cn/lambda}.
\newblock Apr., 2019.

\bibitem{googlefunction}
\url{cloud.google.com/functions}.
\newblock Apr., 2019.

\bibitem{azurefunction}
\url{azure.microsoft.com/en-us/services/functions}.
\newblock Apr., 2019.

\bibitem{Alicloud}
\url{https://cn.aliyun.com/product/fc}.
\newblock Apr., 2020.

\bibitem{openwhisk}
\url{openwhisk.apache.org}.
\newblock Apr., 2019.

\bibitem{serverless_report}
\url{serverless.com/blog/2018-serverless-community-survey-huge-growth-usage}.
\newblock Apr., 2019.

\bibitem{CRIU}
\url{https://github.com/checkpoint-restore/criu}.
\newblock Apr., 2019.

\bibitem{openwhisk_doc}
\url{github.com/apache/openwhisk/blob/90c20a847b9a70b43e316fd89a0a15ae2ee39cc4/docs/annotations.md}.
\newblock Apr., 2019.

\bibitem{openwhisk_pydoc}
\url{https://github.com/apache/openwhisk/blob/master/docs/actions-python.md}.
\newblock Apr., 2019.

\bibitem{openwhisk_dockerdoc}
\url{https://github.com/apache/openwhisk/blob/master/docs/actions-docker.md}.
\newblock Apr., 2019.

\bibitem{openwhisk_doc_main}
\url{https://github.com/apache/openwhisk/blob/master/docs/actions-python.md}.
\newblock Apr., 2019.

\bibitem{Firecracker}
Firecracker lightweight virtualization for serverless computing.
\newblock \url{https://aws.amazon.com/
  blogs/aws/firecracker-lightweight-virtualization-for-serverless-computing/}.
\newblock Apr., 2019.

\bibitem{Knative}
Knative.
\newblock \url{ https://github.com/knative}.
\newblock Apr., 2019.

\bibitem{gVisor}
Open-sourcing gvisor, a sandboxed container runtime.
\newblock \url{https://cloud.google.com/blog/
  products/gcp/open-sourcing-gvisor-a-sandboxed-container-runtime. }.
\newblock Apr., 2019.

\bibitem{akkus2018sand}
Istemi~Ekin Akkus, Ruichuan Chen, Ivica Rimac, Manuel Stein, and Klaus Satzke.
\newblock {SAND}: Towards high-performance serverless computing.
\newblock In {\em ATC}, pages 923--935, 2018.

\bibitem{baldini2017serverless}
Ioana Baldini, Paul Castro, Kerry Chang, Perry Cheng, Stephen Fink, Vatche
  Ishakian, Nick Mitchell, Vinod Muthusamy, Rodric Rabbah, Aleksander
  Slominski, et~al.
\newblock Serverless computing: Current trends and open problems.
\newblock In {\em Research Advances in Cloud Computing}, pages 1--20. Springer,
  2017.

\bibitem{barroso2003web}
Luiz~Andr{\'e} Barroso, Jeffrey Dean, and Urs H{\"o}lzle.
\newblock Web search for a planet: The google cluster architecture.
\newblock {\em IEEE micro}, (2):22--28, 2003.

\bibitem{DBLP:conf/cloud/Brewer15}
Eric~A. Brewer.
\newblock Kubernetes and the path to cloud native.
\newblock In {\em Proceedings of the Sixth {ACM} Symposium on Cloud Computing,
  SoCC 2015, Kohala Coast, Hawaii, USA, August 27-29, 2015}, page 167, 2015.

\bibitem{dean2013tail}
Jeffrey Dean and Luiz~Andr{\'e} Barroso.
\newblock The tail at scale.
\newblock {\em Communications of the ACM}, 56(2):74--80, 2013.

\bibitem{DBLP:conf/iwqos/0001YLCXL13}
Xin Dong, Jiadi Yu, Yuan Luo, Yingying Chen, Guangtao Xue, and Minglu Li.
\newblock Achieving secure and efficient data collaboration in cloud computing.
\newblock In {\em 21st {IEEE/ACM} International Symposium on Quality of
  Service, IWQoS 2013, Montreal, Canada, 3-4 June 2013}, pages 195--200.
  {IEEE}, 2013.

\bibitem{du2020catalyzer}
Dong Du, Tianyi Yu, Yubin Xia, Binyu Zang, Guanglu Yan, Chenggang Qin, Qixuan
  Wu, and Haibo Chen.
\newblock Catalyzer: Sub-millisecond startup for serverless computing with
  initialization-less booting.
\newblock In {\em Proceedings of the Twenty-Fifth International Conference on
  Architectural Support for Programming Languages and Operating Systems}, pages
  467--481, 2020.

\bibitem{gautam2012analysis}
Natarajan Gautam.
\newblock {\em Analysis of queues: Methods and applications}.
\newblock CRC Press, 2012.

\bibitem{DBLP:conf/ccs/GoyalPSW06}
Vipul Goyal, Omkant Pandey, Amit Sahai, and Brent Waters.
\newblock Attribute-based encryption for fine-grained access control of
  encrypted data.
\newblock In Ari Juels, Rebecca~N. Wright, and Sabrina De~Capitani
  di~Vimercati, editors, {\em Proceedings of the 13th {ACM} Conference on
  Computer and Communications Security, {CCS} 2006, Alexandria, VA, USA,
  Ioctober 30 - November 3, 2006}, pages 89--98. {ACM}, 2006.

\bibitem{DBLP:conf/fast/HarterSLAA16}
Tyler Harter, Brandon Salmon, Rose Liu, Andrea~C. Arpaci{-}Dusseau, and
  Remzi~H. Arpaci{-}Dusseau.
\newblock Slacker: Fast distribution with lazy docker containers.
\newblock In {\em 14th {USENIX} Conference on File and Storage Technologies,
  {FAST} 2016, Santa Clara, CA, USA, February 22-25, 2016}, pages 181--195,
  2016.

\bibitem{DBLP:journals/usenix-login/HendricksonSOHV16}
Scott Hendrickson, Stephen Sturdevant, Edward Oakes, Tyler Harter,
  Venkateshwaran Venkataramani, Andrea~C. Arpaci{-}Dusseau, and Remzi~H.
  Arpaci{-}Dusseau.
\newblock Serverless computation with openlambda.
\newblock {\em login Usenix Mag.}, 41(4), 2016.

\bibitem{jonas2019cloud}
Eric Jonas, Johann Schleier-Smith, Vikram Sreekanti, Chia-Che Tsai, Anurag
  Khandelwal, Qifan Pu, Vaishaal Shankar, Joao Carreira, Karl Krauth, Neeraja
  Yadwadkar, et~al.
\newblock Cloud programming simplified: a berkeley view on serverless
  computing.
\newblock {\em arXiv preprint arXiv:1902.03383}, 2019.

\bibitem{functionbench}
Jeongchul Kim and Kyungyong Lee.
\newblock Functionbench: A suite of workloads for serverless cloud function
  service.
\newblock In {\em CLOUD}, pages 502--504. IEEE, 2019.

\bibitem{DBLP:conf/hotos/KollerW17}
Ricardo Koller and Dan Williams.
\newblock Will serverless end the dominance of linux in the cloud?
\newblock In {\em Proceedings of the 16th Workshop on Hot Topics in Operating
  Systems, HotOS 2017, Whistler, BC, Canada, May 8-10, 2017}, pages 169--173,
  2017.

\bibitem{DBLP:journals/cacm/MadhavapeddyS14}
Anil Madhavapeddy and David~J. Scott.
\newblock Unikernels: the rise of the virtual library operating system.
\newblock {\em Commun. {ACM}}, 57(1):61--69, 2014.

\bibitem{DBLP:conf/sosp/MancoLSMKSYRH17}
Filipe Manco, Costin Lupu, Florian Schmidt, Jose Mendes, Simon Kuenzer, Sumit
  Sati, Kenichi Yasukata, Costin Raiciu, and Felipe Huici.
\newblock My {VM} is lighter (and safer) than your container.
\newblock In {\em Proceedings of the 26th Symposium on Operating Systems
  Principles, Shanghai, China, October 28-31, 2017}, pages 218--233, 2017.

\bibitem{DBLP:conf/icdcsw/McGrathB17}
M.~Garrett McGrath and Paul~R. Brenner.
\newblock Serverless computing: Design, implementation, and performance.
\newblock In {\em 37th {IEEE} International Conference on Distributed Computing
  Systems Workshops, {ICDCS} Workshops 2017, Atlanta, GA, USA, June 5-8, 2017},
  pages 405--410, 2017.

\bibitem{sock}
Edward Oakes, Leon Yang, Dennis Zhou, Kevin Houck, Tyler Caraza{-}Harter,
  Andrea~C. Arpaci{-}Dusseau, and Remzi~H. Arpaci{-}Dusseau.
\newblock {SOCK:} serverless-optimized containers.
\newblock {\em login Usenix Mag.}, 43(3), 2018.

\bibitem{DBLP:conf/nsdi/PuVS19}
Qifan Pu, Shivaram Venkataraman, and Ion Stoica.
\newblock Shuffling, fast and slow: Scalable analytics on serverless
  infrastructure.
\newblock In {\em 16th {USENIX} Symposium on Networked Systems Design and
  Implementation, {NSDI} 2019, Boston, MA, February 26-28, 2019}, pages
  193--206, 2019.

\bibitem{DBLP:conf/micro/ShahradBW19}
Mohammad Shahrad, Jonathan Balkind, and David Wentzlaff.
\newblock Architectural implications of function-as-a-service computing.
\newblock In {\em Proceedings of the 52nd Annual {IEEE/ACM} International
  Symposium on Microarchitecture, {MICRO} 2019, Columbus, OH, USA, October
  12-16, 2019}, pages 1063--1075, 2019.

\bibitem{DBLP:journals/corr/abs-1810-09679}
Vaishaal Shankar, Karl Krauth, Qifan Pu, Eric Jonas, Shivaram Venkataraman, Ion
  Stoica, Benjamin Recht, and Jonathan Ragan{-}Kelley.
\newblock numpywren: serverless linear algebra.
\newblock {\em CoRR}, abs/1810.09679, 2018.

\bibitem{DBLP:conf/asplos/ShenSSBDRW19}
Zhiming Shen, Zhen Sun, Gur{-}Eyal Sela, Eugene Bagdasaryan, Christina
  Delimitrou, Robbert van Renesse, and Hakim Weatherspoon.
\newblock X-containers: Breaking down barriers to improve performance and
  isolation of cloud-native containers.
\newblock In Iris Bahar, Maurice Herlihy, Emmett Witchel, and Alvin~R. Lebeck,
  editors, {\em Proceedings of the Twenty-Fourth International Conference on
  Architectural Support for Programming Languages and Operating Systems,
  {ASPLOS} 2019, Providence, RI, USA, April 13-17, 2019}, pages 121--135.
  {ACM}, 2019.

\bibitem{DBLP:journals/ijnsec/TaiCH20}
Wei{-}Liang Tai, Ya{-}Fen Chang, and Wen{-}Hsin Huang.
\newblock Security analyses of a data collaboration scheme with hierarchical
  attribute-based encryption in cloud computing.
\newblock {\em I. J. Network Security}, 22(2):212--217, 2020.

\bibitem{DBLP:conf/usenix/ThalheimBFK18}
J{\"{o}}rg Thalheim, Pramod Bhatotia, Pedro Fonseca, and Baris Kasikci.
\newblock Cntr: Lightweight {OS} containers.
\newblock In {\em 2018 {USENIX} Annual Technical Conference, {USENIX} {ATC}
  2018, Boston, MA, USA, July 11-13, 2018}, pages 199--212, 2018.

\bibitem{DBLP:conf/pppj/2018}
Eli Tilevich and Hanspeter M{\"{o}}ssenb{\"{o}}ck, editors.
\newblock {\em Proceedings of the 15th International Conference on Managed
  Languages {\&} Runtimes, ManLang 2018, Linz, Austria, September 12-14, 2018}.
  {ACM}, 2018.

\bibitem{Venkatesh2019fast}
Ranjan~Sarpangala Venkatesh, Till Smejkal, Dejan~S. Milojicic, and Ada
  Gavrilovska.
\newblock Fast in-memory criu for docker containers.
\newblock In {\em Proceedings of the International Symposium on Memory
  Systems}, MEMSYS ¡¯19, pages 53--65, New York, NY, USA, 2019. Association
  for Computing Machinery.

\bibitem{DBLP:conf/sosp/VrableMCMVSVS05}
Michael Vrable, Justin Ma, Jay Chen, David Moore, Erik Vandekieft, Alex~C.
  Snoeren, Geoffrey~M. Voelker, and Stefan Savage.
\newblock Scalability, fidelity, and containment in the potemkin virtual
  honeyfarm.
\newblock In {\em Proceedings of the 20th {ACM} Symposium on Operating Systems
  Principles 2005, {SOSP} 2005, Brighton, UK, October 23-26, 2005}, pages
  148--162, 2005.

\bibitem{DBLP:conf/eurosys/WangHW19}
Kai{-}Ting~Amy Wang, Rayson Ho, and Peng Wu.
\newblock Replayable execution optimized for page sharing for a managed runtime
  environment.
\newblock In {\em Proceedings of the Fourteenth EuroSys Conference 2019,
  Dresden, Germany, March 25-28, 2019}, pages 39:1--39:16, 2019.

\bibitem{wang2018peeking}
Liang Wang, Mengyuan Li, Yinqian Zhang, Thomas Ristenpart, and Michael Swift.
\newblock Peeking behind the curtains of serverless platforms.
\newblock In {\em ATC}, pages 133--146, 2018.

\end{thebibliography}

\end{document}